\documentclass[12pt]{article}
\usepackage{indentfirst}
\usepackage{graphicx}
\usepackage{placeins}
\usepackage{slashed}
\usepackage{gensymb}
\usepackage{amsmath}
\usepackage{amssymb}
\usepackage{feynmp-auto}
\usepackage{hyperref}
\newcommand\pubnumber{}
\newcommand\pubdate{\today}

\def\napoli{Bethe Center for Theoretical Physics\\
Bonn University, 53115 Bonn, GERMANY}
\def\support{\footnote{zhongyi@th.physik.uni-bonn.de}}

\def\Title#1{\begin{center} {\Large #1 } \end{center}}
\def\Author#1{\begin{center}{ \sc #1} \end{center}}
\def\Address#1{\begin{center}{ \it #1} \end{center}}

\newcommand\pubblock{\rightline{\begin{tabular}{l} \pubnumber\\
         \pubdate  \end{tabular}}}
\newenvironment{Abstract}{\begin{quotation}  }{\end{quotation}}

\def\Acknowledgements{\bigskip  \bigskip \begin{center} \begin{large}
             \bf ACKNOWLEDGEMENTS \end{large}\end{center}}
\begin{document}

\def\gsim{\:\raisebox{-0.5ex}{$\stackrel{\textstyle>}{\sim}$}\:}
\def\lsim{\:\raisebox{-0.5ex}{$\stackrel{\textstyle<}{\sim}$}\:}

\begin{titlepage}
\pubblock

%\vfill
\Title{Constraints on a Light Leptophobic Mediator from LEP Data}
\Author{Manuel Drees, Zhongyi Zhang\support}
\Address{\napoli}

\begin{Abstract}
  We apply data taken at the $e^+e^-$ collider LEP in the 1990's at
  center--of--mass energy up to 209 GeV to constrain Dark Matter
  models with a light leptophobic spin$-1$ mediator $R$.  We assume
  that the dark sector particle (DSP) is a spin$-1/2$ fermion
  $\chi$. This scenario is well studied in the context of LHC searches
  for mediator mass from 100 GeV to several TeV. Emission of the mediator
  off a quark or antiquark at LEP gives rise to di--jet plus missing
  energy and $4-$jet signatures, which we use to limit the relevant
  couplings. We focus on scenarios with $2 m_\chi > m_R$, which are
  poorly constrained by LHC data. We recast published searches by the
  ALEPH collaboration. For $m_\chi \lsim 20$ GeV the best bounds
  result from an analysis at $\sqrt{s} \simeq M_Z$ of di--jet plus
  missing energy events. For heavier DSP but $m_R \lsim 70$ GeV
  meaningful bounds can be derived from a four jet analysis at
  $\sqrt{s} = 183$ GeV. Unfortunately published searches using four
  jet final states at $\sqrt{s} \simeq M_Z$ use only a small fraction
  of the total data sample. Moreover, all published searches for
  di--jet plus missing energy final states at $\sqrt{s} \geq 130$ GeV
  have poor efficiency for our model; we therefore design new cuts
  that combine good background rejection with higher efficiency.
  Re--analyzing the higher energy data using our new cuts, and an
  analysis of the complete four jet data sample taken at
  $\sqrt{s}\simeq M_Z$, can explore new regions of parameter space.
\end{Abstract}

\end{titlepage}
\def\thefootnote{\fnsymbol{footnote}}
\setcounter{footnote}{0}

\section{Introduction}

The Standard Model (SM) of particle physics lacks an adequate
candidate for dark matter \cite{feng2010dark}. Particle physics
explanations of dark matter therefore typically include a ``dark
sector'' containing (at least) one stable dark matter candidate
(called DSP for Dark Sector Particle in this paper), and at least one
mediator coupling the DSP to SM particles. In principle this mediator
could be the well--known $Z$ or $125$ GeV Higgs boson, but these
``portal'' models are by now very tightly constrained
\cite{Banerjee:2016hsk, Hoferichter:2017olk, Athron:2017kgt}. Here we
are interested in models where the mediator is not part of the SM.
Frequently it is a massive scalar or vector boson. Therefore, a
simplified model approach \cite{DiFranzo:2013vra, Abdallah:2015ter,
  Boveia:2016mrp} allows to constrain many UV complete extensions of
the SM. Simplified models usually have a relatively small number of
free parameters, allowing exhaustive scans of the parameter
space. Models designed to describe the scattering of DSPs on ordinary
matter, as in ``direct search'' experiments, have to specify the
couplings of the mediator(s) to hadrons and to the DSP. This suffices
to fix the rate of monojet (and similar) events at the LHC. Since no
excess of such events has been found, LHC data have given strong
constraints for mediator masses below about $1$ TeV that can decay
invisibly, e.g. into a pair of dark matter particles
\cite{Aaboud:2017phn,Sirunyan:2017hci}. Moreover, mediator masses
roughly between $1$ and $2.5$ TeV are also constrained by searches for
di--jet final states \cite{Sirunyan:2016iap,Aaboud:2017yvp}. Very
recently this range has been extended downward by using special search
strategies \cite{Aaboud:2018fzt}; preliminary results using events
with a hard third jet recoiling against a ``fat jet'' allowed CMS to
extend the search range down to $50$ GeV \cite{Bauce:2017jir}, for
coupling strength to (light) quarks $\gsim 0.2$.\footnote{This
  analysis also found a slight excess of events corresponding to a
  mediator mass of about 115 GeV.}

It should be noted that more complete models are often subject to
additional constraints. For example, $Z^\prime$ models based on
extending the SM gauge group with an additional $U(1)$ factor were
investigated in \cite{Babu:1997st, Dudas:2009uq, Accomando:2010fz,
  Fox:2011qd, Frandsen:2011cg, Alves:2013tqa}, and supersymmetric models in
\cite{jungman1996supersymmetric, Kvellestad:2017rwl,
  Bagnaschi:2017tru}. However, many or most of these constraints are
not directly related to the dark matter problem.

Although the Large Electron Positron collider (LEP) at CERN ceased
operations nearly twenty years ago, and only covered center--of--mass
(cms) energies up to $209$ GeV, the cleaner environment and the
distinct energy range still offer some advantages for certain regions
of parameter space. In this work we focus on a simplified model with a
vector mediator $R$ coupling to the DSP and quarks. We use the
framework of ref.\cite{Frandsen:2012rk}, which starts from the very
general assumption that the new mediator couples to all different
kinds of SM particles, including gauge bosons, leptons and quarks. It
uses LHC data (from run $1$) in order to derive stringent upper bounds
on many of these couplings. These constraints are quite strong if
on--shell decays of the mediator to DSPs are possible, or if the
mediator has sizable couplings to leptons.  We saw above that (in some
cases still preliminary) constraints from LHC searches for purely
hadronic final states have become quite strong, if $m_R > 50$
GeV. However, the published constraints apply to couplings to first
generation quarks, which are strongly constrained by direct dark
matter searches. The annihilation cross section of the DSP $\chi$ into
hadronic final states can therefore still be sufficiently large for
$\chi$ to be a good thermal WIMP (Weakly Interacting Massive Particle)
candidate \cite{Kolb:1990vq} in standard cosmology. In this study we
also focus on the $\bar{\chi}\chi R$ and $\bar{q}qR$ couplings. Note
that together with the masses $m_R$ and $m_\chi$ these are the key
parameters determining both the direct WIMP detection rate and (if the
other couplings are small) the relic density.

There are also purely theoretical constraints on the model.
Ref.\cite{Frandsen:2012rk} derived an upper bound on the couplings
from the requirement that perturbation theory can be applied, since we
do not know how to constrain these couplings otherwise. Moreover, as
pointed out in ref.\cite{Kahlhoefer:2015bea}, if $R$ has
non--vanishing axial vector couplings to $\chi$, unitarity imposes an
upper bound on the ratio of DSP and mediator masses. We apply the same
perturbativity and unitarity conditions in the part of parameter space
that could have been probed by LEP experiments.

LHC data only probe configurations where the mediator is essentially
on--shell. In contrast, in this study, which focuses on a light
mediator, we consider cases where the DSP pair can only be generated
through off--shell processes. We notice an enhancement of the cross
section if the mediator has an axial vector coupling to $b-$quarks and
$m_R < m_b$ because the longitudinal part of the mediator contributes
a term $\propto m_b^2 / m_R^2$ to the four--jet cross
section. Similarly, in the presence of an axial vector coupling to the
DSP the di--jet plus missing energy cross section may increase with
increasing $m_\chi$, contrary to naive expectations. However, the
unitarity constraints imply that these terms cannot be arbitrarily
large.

The remainder of this article is organized as follows. In section
\ref{sec:2} the Lagrangian of the simplified model is introduced, and
bounds on the relevant couplings from theoretical considerations and
non--collider experiments are discussed. In section \ref{sec:3} we
recast searches for di--jet plus missing energy and $4-$jet final
states performed by the ALEPH collaboration \cite{Decamp:1990jra,
  Buskulic:1994wz}. We discuss the bounds resulting from these
published searches and the cut efficiencies when applied to our
model. In section \ref{sec:4} we introduce a set of specially designed
cuts for the di--jet plus missing energy signature that have much
higher efficiency for our signal than the published searches. Although
we do not include the detector simulation in the test of the
background suppression, the result still shows the potential of the
LEP data to improve on the bounds derived in section
\ref{sec:3}. Finally, section \ref{sec:5} is devoted to a summary and
some conclusions.

\section{The Simplified Model}
\label{sec:2}

In this section we first describe the Lagrangian of the simplified
model we consider. We then discuss limits on the model parameters that
follow if the DSP is assumed to be a thermal WIMP, which is subject to
stringent constraints from direct dark matter search experiments. In
the following two subsections we discuss upper bounds on the couplings
that result from perturbativity and unitarity constraints. In the
final subsection the pre--collider bounds on the remaining free
parameters are summarized and our final choice of free parameters is
discussed.

\subsection{Lagrangian and Free Parameters}

As discussed in the Introduction, we consider a simplified model
\cite{Frandsen:2012rk} where a massive spin$-1$ mediator connects the
DSP to SM particles. The Lagrangian can then be written as
\begin{equation} \label{L_tot}
\mathcal{L} = \mathcal{L}_{\rm SM} + \mathcal{L}_{\rm DSP}
+ \mathcal{L}_R + \mathcal{L}_I\,.
\end{equation}
We assume the DSP to be a spin$-1/2$ Dirac fermion. A Majorana fermion
cannot have a vector interaction, but is otherwise basically the same
as a Dirac fermion for our purposes.\footnote{A complex scalar DSP
  behaves similar to a Majorana DSP if $m_R > m_b, m_\chi$. However,
  the contribution from the exchange of longitudinal messenger
  particles vanishes identically in this case, i.e. there are no terms
  that are enhanced by $m_b m_\chi / m_R^2$.} The DSP part of the
Lagrangian is therefore:
\begin{equation} \label{L_DSP}
\mathcal{L}_{\rm DSP} = \bar{\chi}(i\slashed{\partial} - m_\chi) \chi \,.
\end{equation}
In MadGraph convention \cite{alwall2011madgraph} the mediator part of
the Lagrangian is:
\begin{equation} \label{L_R}
\mathcal{L}_R  = -\frac{1}{4} F^{\mu\nu} F_{\mu\nu} -
\frac{1}{2} m_R^2 R^\mu R_\mu\,, \ \ \ {\rm with} \ F_{\mu\nu} \equiv
\partial_\mu R_\nu - \partial_\nu R_\mu\,.
\end{equation}
Finally, the interactions of the mediator with fermions are described by the
Lagrangian
\begin{equation} \label{L_I}
\mathcal{L}_I = \sum_q R_\mu \bar{q} \gamma^\mu \left( g^V_q
- g^A_q \gamma^5 \right) q + R_\mu \bar{\chi} \gamma^\mu \left( g^V_\chi
- g^A_\chi \gamma^5 \right) \chi \,.
\end{equation}

The free parameters of our model are thus the mediator mass $m_R$, the
DSP mass $m_\chi$, and the couplings of the mediator to quarks
($g_q^V$, $g_q^A$) and to the DSP ($g_\chi^V$, $g_\chi^A$). In total,
there are $16$ parameters. However, since this study uses data from
$e^+ e^-$ collision up to $\sqrt{s} = 209$ GeV, top quarks cannot
contribute to the final state. Therefore the couplings $g^V_t$ and
$g^A_t$ are irrelevant, so that $14$ relevant free parameters remain.

An exhaustive scan of a $14-$dimensional parameter space is not
feasible with our computational resource. However, as we will see
in the following subsections, non--collider constraints force
many of these couplings to be very small, so that we can set them to
zero for our purposes.

\subsection{Dark Matter Constraints}

In the standard thermal WIMP scenario, the dark matter relic density
is essentially inversely proportional to the total DSP annihilation
cross section computed in the non--relativistic limit
\cite{Kolb:1990vq}. In our model the DSP can always annihilate into
sufficiently light quarks, with cross section \cite{Chala:2015ama}:
\begin{eqnarray} \label{ann_qq}
v\sigma( \bar{\chi} \chi & \rightarrow & \bar{q}q ) \simeq \frac {3m_\chi^2}
{2 \pi (m_R^2 - 4m_\chi^2 )^2} \sqrt { 1 - \frac {m_q^2} {m_\chi^2}} \\
&\cdot & \left[ (g_q^V)^2 (g_\chi^V)^2 \left( 2 + \frac {m_q^2}{m_\chi^2}
\right) + 2 (g_q^A)^2 (g_\chi^V)^2 \left( 1 - \frac {m_q^2}{m_\chi^2} \right)
+ (g_q^A)^2 (g_\chi^A)^2 \frac {m_q^2} {m_\chi^2} \frac {(4m_\chi^2-m_R^2)^2}
{m_R^4} \right]\,. \nonumber
\end{eqnarray}
Here $v$ is the relative velocity between $\chi$ and $\bar\chi$.  The
last term on the right--hand side (rhs) of eq.(\ref{ann_qq}) is due to
the exchange of longitudinal $R-$bosons. Note that is is {\em
  enhanced} $\propto m_\chi^2 m_q^2 / m_R^4$ for small mediator
masses; at the same time it is {\em suppressed}
$\propto m_q^2 / m_\chi^2$ if $m_R > 2 m_\chi \gg m_q$. The numerator
of this term implies that it does not have a pole at
$s \simeq 4 m_\chi^2 = m_R^2$. If the vectorial couplings do not
vanish, this term is therefore only relevant if the exchanged mediator
is quite far off--shell. Notice also that this term is proportional to
the product of axial vector couplings, i.e. it is absent for a purely
vectorial theory. At the same time it is the only term that survives
for vanishing vector couplings, e.g. if $\chi$ is a Majorana particle.

Moreover, for $m_\chi > m_R$ a $\chi \bar\chi$ pair can also annihilate
into two mediators, which subsequently decay to quarks. The corresponding
cross section is \cite{Chala:2015ama}:
\begin{eqnarray} \label{ann_RR}
v\sigma( \bar{\chi} \chi \rightarrow RR) &=& \frac{ (m_\chi^2 - m_R^2)^{3/2} }
{4\pi m_\chi (m_R^2 - 2m_\chi^2)^2 } \\
&\cdot& \left\{ 8 (g^A_\chi)^2 (g^V_\chi)^2 \frac {m_\chi^2} {m_R^2}
+ \left[ (g^A_\chi)^4 - 6(g^A_\chi)^2 (g^V_\chi)^2 + (g^V_\chi)^4 \right]
 \right\} \,. \nonumber
\end{eqnarray}
The first term in the second line again gives an enhancement
$\propto m_\chi^2/m_R^2$. Note that in the limit $v \rightarrow 0$,
which we applied here, the contribution
$\propto (g^A_\chi)^4 / m_R^4$, which is due to the production of two
longitudinal $R$ bosons, vanishes. Moreover, the cross section
(\ref{ann_RR}) is quite strongly phase space suppressed near threshold
where $m_\chi \simeq m_R$.

Since the predicted DSP relic density is inversely proportional to the
total $\chi \bar\chi$ annihilation cross section, requiring that the
predicted DSP density is not larger than the total observed dark
matter density imposes a lower bound on (sums of products of) the
relevant couplings if the masses are fixed. The detailed analysis of
ref.\cite{Chala:2015ama} shows that for $m_R \leq 100$ GeV this bound
is easily satisfied if all axial vector couplings are $\gsim 0.3$ even
for vanishing vector couplings. We will see below that LEP data only
allow to probe significantly smaller $m_R$. We confirm that for
coupling strengths of interest to LEP physics, in standard cosmology
the thermal DSP relic density is always much below the desired dark
matter density, unless the DSP is very light (with $m_\chi < m_R$ so
that $\chi \bar \chi \rightarrow R R$ annihilation is suppressed) and
has very small couplings to light quarks (see below).

The signal in direct dark matter detection experiments depends
essentially on the mass of the dark matter particle and its
scattering cross section on nucleons. For the latter one usually
distinguishes between spin--dependent (SD) and spin--independent (SI)
contributions. The corresponding cross sections can be written as
\cite{Chala:2015ama}:
\begin{equation} \label{sig_dir}
\sigma_N^{\rm SD} = a_N^2 \frac {3\mu^2_N} {\pi m_R^4}\,; \ \
\sigma_N^{\rm SI} = f_N^2 \frac{3\mu^2_N}{\pi m_R^4}\,.
\end{equation}
Here $N=n,p$ and
\begin{equation} \label{mu}
\mu_N = \frac {m_\chi m_N} {m_\chi+m_N}
\end{equation}
is the reduced mass of the DSP--nucleon system. The coefficients
$f_N$ appearing in $\sigma_N^{\rm SI}$ are simply given by products of
couplings:
\begin{equation} \label{fN}
f_p = g_\chi^V ( 2g_u^V + g_d^V )\,;\ \ f_n = g_\chi^V (g_u^V + 2g_d^V)\,,
\end{equation}
where the differences are due to the different valence quark content
of neutrons and protons. Note that sea quarks do not contribute, since
quarks and antiquarks couple with opposite sign to $R$; their
contributions cancel, since here the coherent coupling to the entire
nucleon (in fact, in most cases to an entire nucleus) is
relevant. Finally, the coefficients $a_N$ appearing in
$\sigma_N^{\rm SD}$ are:
\begin{equation} \label{aN}
a_N = g^A_\chi \sum_{q=u,d,s} \Delta q^{(N)} g_q^A\,.
\end{equation}
Here $\Delta q^{(N)}$ is the contribution of the spin of quark $q$ to
the total spin of nucleon $N$. They can be determined from polarized
deep--inelastic scattering experiments. The current
Particle Data Group values \cite{Olive:2016xmw} are:
\begin{eqnarray} \label{Delta}
\Delta u^{(p)} &=& \Delta d^{(n)} = 0.84 \pm 0.02\,; \nonumber  \\
\Delta u^{(n)} &=& \Delta d^{(p)} = -0.43 \pm 0.02\,; \\
\Delta s^{(p)} &=& \Delta s^{(n)} = -0.09 \pm 0.02\,. \nonumber
\end{eqnarray}

There are strong upper bounds on the spin--independent scattering
cross section on the proton. For $m_\chi \gsim 5$ GeV the tightest
constraint comes from the PandaX--II \cite{Tan:2016zwf} experiment,
whereas CRESST \cite{Angloher:2015ewa} data impose significant
constraints for $m_\chi \gsim 0.5$ GeV. We will see below that LEP
data can only probe scenarios with $m_R < 100$ GeV. These bounds
require $g^V_{u,d}$ to be below $0.1$, usually much below this
value. Such small couplings have little influence on LEP physics, so
we set $g^V_u = g^V_d = 0$.\footnote{The bounds on the
  spin--independent cross section have been derived under the
  assumption of equal scattering cross section on neutrons and
  protons, which need not be the case in our scenario. In fact, the
  cross section for scattering on any one isotope can be made to
  vanish for a particular (negative) ratio of $g_u^V/g_d^V$. However,
  by now experiments using many different isotopes have been
  performed, allowing to constrain $g_u^V$ and $g_d^V$ separately.}

The upper bounds on the spin--dependent cross sections become quite
weak for WIMP mass below 4 GeV, but the bound on $\sigma_n^{\rm SD}$
is still below $10^{-2}$ pb for $m_\chi = 5$ GeV
\cite{Amole:2017dex}. If $m_R \leq 10$ GeV this constraint suffices to
imply $g_{u,d}^A \leq 0.1$, the bound on $g^A_s$ being somewhat weaker
but still strong enough to force these couplings to be negligible for
LEP physics. On the other hand, for $m_R \geq 50$ GeV ${\cal O}(1)$
axial vector couplings are allowed even for the light quarks if we scale
the bound on the scattering cross section by the ratio of the predicted
$\chi$ relic density and the total observed dark matter density. However,
in that case $\chi$ does not make a good thermal dark matter candidate. In
most scenarios where the predicted $\chi$ relic density in standard cosmology
is at least a sizable fraction of the observed dark matter density
the upper bound on the spin dependent cross section for $m_\chi \gsim 4$
GeV requires the axial vector couplings to be too small to significantly
affect LEP cross section. We therefore set $g^A_u=g^A_d = g^A_s = g^A_c = 0$;
we require vanishing axial vector coupling to charm quarks since strange
and charm quarks reside in the same $SU(2)$ doublet.

We are then left with eight free parameters: four couplings of $R$ to
quarks, two couplings of $R$ to the DSP, and the masses of $R$ and the
DSP.

\subsection{Perturbativity Condition}

We will use leading order perturbation theory to derive constraints on
our model from published LEP data. Perturbation theory becomes
unreliable when the couplings become too large. Our calculations
depend on the SM electroweak couplings, which are perturbative, and on
the couplings of the mediator $R$. We constrain the latter through the
simple condition
\begin{equation} \label{pert1}
\Gamma_R < m_R\,,
\end{equation}
where $\Gamma_R$ is the total decay width of $R$. $R$ can decay into
$q \bar q$ and $\chi \bar \chi$ pairs, with partial widths:
\begin{eqnarray} \label{Gamma_R}
\Gamma(R \rightarrow q\bar{q} ) &=& \frac {m_R} {4\pi} \sqrt{ 1 - 4z_q }
\left[ (g^V_q)^2 + (g^A_q)^2 + z_q \left( 2(g^V_q)^2 - 4(g^A_q)^2 \right)
 \right]\,; \nonumber \\
\Gamma(R \rightarrow \chi \bar{\chi}) &=& \frac {m_R} {12\pi}
\sqrt{ 1 - 4z_\chi } \left[ (g^V_\chi)^2 + (g^A_\chi)^2 +z_\chi
\left( 2(g^V_\chi)^2 - 4(g^A_\chi)^2 \right) \right]\,.
\end{eqnarray}
Here $z_f \equiv m_f^2/m_R^2$. The factor of $3$ in the first equation
comes from the colors of quarks. Of course, these widths are nonzero
only for $m_R > 2 m_f$, i.e. $z_F < 0.25$. The perturbativity condition
can thus be written as
\begin{equation} \label{pert2}
\sum_{2m_f < m_R} N_f \sqrt{1-4z_f} \left[ (g^V_f)^2 + (g^A_f)^2 + z_f
\left( 2 (g^V_f)^2 - 4 (g^A_f)^2 \right) \right] < 12\pi\,.
\end{equation}
This constraint can be used for $m_R \geq 1$ GeV, so that at least
decays into strange quarks are possible. For somewhat heavier
mediators, which can also decay into $c \bar c$ and perhaps
$\chi \bar \chi$ pairs, the constraint (\ref{pert2}) becomes
stronger. We will only use combinations of parameters that respect
this bound.

\subsection{Unitarity Condition}

Another important kind of constraint has first been discussed in
ref.\cite{Kahlhoefer:2015bea}: unitarity limits the size of the axial
vector couplings of fermions $f$ to the mediator $R$. One way to see
this is to consider the cross section for
$f \bar f \rightarrow R_L R_L$, where $R_L$ denotes a longitudinally
polarized $R$ boson. For fixed (nonzero) relative velocity between
$f$ and $\bar f$, the matrix element scales like $(g^A_f m_f / m_R)^2$.
This violates unitarity, unless
\begin{equation} \label{unitarity}
g_f^A \frac {m_f} {m_R} \leqslant \sqrt{ \frac {\pi} {2} }\,.
\end{equation}
Note that this bound applies both to the DSP, $f=\chi$, and to the quarks
with non--vanishing axial vector coupling, $f=q$.

Another derivation of the unitarity constraint starts from the
observation that in a renormalizable theory, $R$ must be a gauge
boson. If fermion $f$ has non--vanishing axial coupling $g^A_f$ to
$R$, the two--component fermions $f_L$ and $f_R$ must transform
differently under the $R$ gauge symmetry. This implies that the
(Dirac) mass term $m_f \overline{f_L} f_R$ is not invariant under the
$R$ gauge symmetry. Hence $m_f$ must be due to the vacuum expectation
value of some Higgs field that carries $R$ charge. The upper bound
(\ref{unitarity}) then follows from the upper bound on the Yukawa
coupling that gives rise to $m_f$.

The bound (\ref{unitarity}) limits the size of the enhancement due to
the exchange of longitudinal $R-$bosons with axial vector coupling to
massive fermions; see the discussion of eq.(\ref{ann_qq}) above. We
will see below that similar terms also appear in our signal cross
sections.  Neglecting the unitarity constraint (\ref{unitarity}) could
thus lead to overly optimistic conclusions regarding the sensitivity
of collider data to our model.

\subsection{Summary: Free Parameters of the Model}

The perturbativity condition (\ref{pert2}) is quite weak. The
unitarity constraint (\ref{unitarity}) can be strong for small $m_R$,
but only applies to the axial vector couplings, and in any case still
allows non--negligible couplings. These constraints therefore do not
reduce the number of free parameters, i.e. we still have the eight
free parameters enumerated at the end of Sec.~2.2. This parameter
space is still too large for a thorough exploration.

We therefore assume equal vector couplings of $s,\, c$ and $b$
quarks. Recall that we set the vector couplings of $u$ and $d$ quarks
to zero in order to satisfy constraints from direct detection
experiments. As mentioned in the Introduction, we will investigate
final states with either two jets and two DSPs, or with four jets. The
searches we will use to probe $q \bar q \chi \bar\chi$ production do
not require any flavor tagging, so to good approximation this cross
section only depends on the sum $(g^V_s)^2 + (g^V_c)^2 + (g^V_b)^2$.
Results for different ratios of the vector couplings therefore can be
derived by simply re--scaling the results presented below. In
contrast, the best published probe of the four--jet final state
requires the detection of at least two $b$ (anti)quarks in the final
state. Since $b \bar b b \bar b$ final states have a significantly
higher probability of satisfying this requirements than final states
with only one $b \bar b$ pair, $g^V_b$ contributes with higher weight
to the final cross section after cuts than $g^V_c$ and $g^V_s$.

Recall that scenarios where a light $R$ can decay into a
$\chi \bar \chi$ pair are strongly constrained by LHC ``monojet''
data. We will thus assume $m_R < 2 m_\chi$. In that case the
(tree--level) cross section for the four--jet final state is
completely independent of the couplings $g^V_\chi$ and
$g^A_\chi$. Moreover, the cross section for $q \bar q \chi \bar\chi$
production is then proportional to the product $(g_q g_\chi)^2$.  It
is thus sufficient to present results for a fixed ratio of the
couplings of the mediator to quarks and to DSPs; results for different
ratios can then be obtained by re--scaling our results presented
below.

In the end we are left with four free parameters: $m_R$, $m_\chi$,
$g_b^A$ and $g^V_q$.

\section{Application of LEP Data}
\label{sec:3}

In this Section we check whether published analyses of LEP data can
impose significant constraints on the parameters of our model.  We
focus on analyses by the ALEPH collaboration
\cite{Decamp:1990jra,Buskulic:1994wz}, because they are based on
well--defined, and clearly described, cuts defining final states that
receive contributions from the two processes we wish to probe. We
expect data from the other three LEP experiments (DELPHI, L3 and OPAL)
to have similar sensitivity, so a combined analysis could lead to
somewhat stronger bounds.

In our numerical analysis we use FeynRules \cite{Alloul:2013bka} to
generate a model file in UFO format \cite{Degrande:2011ua}, MadGraph
\cite{alwall2011madgraph} to simulate the $e^+e^-$ collision, and
Pythia 8.2 \cite{sjostrand2015introduction} to perform the
hadronization. We apply the cuts defining the relevant ALEPH analyses
at the hadron level, neglecting detector resolution effects. In the
following two Subsections we discuss two--jet plus missing energy and
four--jet final states, respectively.

\subsection{Two Jets Plus Missing Energy}

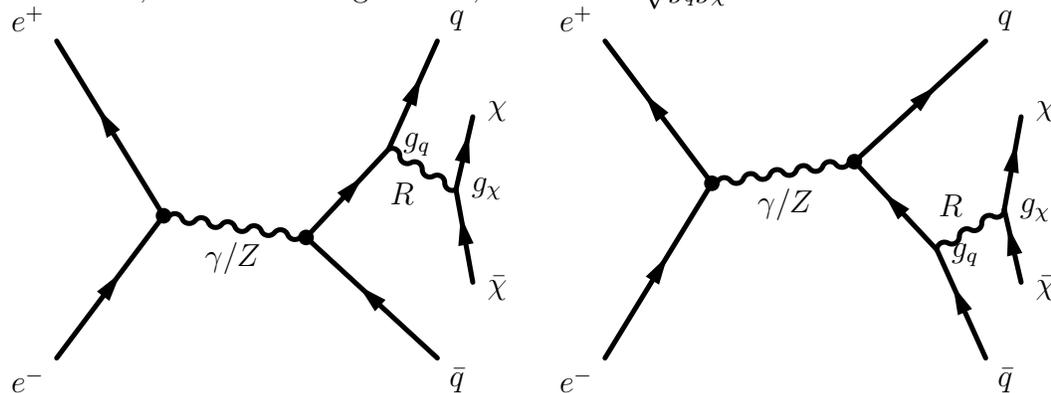
\begin{figure}[thb]
\begin{fmffile}{ee2jjdm}
	\begin{fmfgraph*}(180,120)
	\fmfpen{thick}
	\fmfleft{i1,i2,i3,i4,i5}
	\fmflabel{$e^-$}{i1}
	\fmflabel{$e^+$}{i5}
	\fmfright{o1,o2,o3,o4,o5}
	\fmflabel{$q$}{o5}
	\fmflabel{$\bar{\chi}$}{o2}
	\fmflabel{$\chi$}{o4}
	\fmflabel{$\bar{q}$}{o1}
	\fmf{fermion}{i1,v1,i5}
	\fmf{photon,tension=1.5,label=$\gamma/Z$}{v1,v2}
 	\fmf{fermion}{o1,v2,v3,o5}
	\fmf{photon,tension=0.5,label=$R$}{v3,v4}
	\fmf{fermion}{o2,v4,o4}
	\fmflabel{$g_q$}{v3}
	\fmflabel{$g_\chi$}{v4}
	\fmfdot{v1,v2}
	\end{fmfgraph*}
	\qquad
	\begin{fmfgraph*}(180,120)
	\fmfpen{thick}
	\fmfleft{i1,i2,i3,i4,i5}
	\fmflabel{$e^-$}{i1}
	\fmflabel{$e^+$}{i5}
	\fmfright{o1,o2,o3,o4,o5}
	\fmflabel{$q$}{o5}
	\fmflabel{$\bar{\chi}$}{o2}
	\fmflabel{$\chi$}{o4}
	\fmflabel{$\bar{q}$}{o1}
	\fmf{fermion}{i1,v1,i5}
	\fmf{photon,tension=1.5,label=$\gamma/Z$}{v1,v2}
 	\fmf{fermion}{o1,v3,v2,o5}
	\fmf{photon,tension=0.5,label=$R$}{v3,v4}
	\fmf{fermion}{o2,v4,o4}
	\fmflabel{$g_q$}{v3}
	\fmflabel{$g_\chi$}{v4}
	\fmfdot{v1,v2}
	\end{fmfgraph*}
\end{fmffile}

\caption{Leading order diagrams contributing to the final state (\ref{twojet})
in our model. Note that the mediator $R$ is always off--shell in the region
of parameter space we are interested in.}
\label{fig:diag1}
\end{figure}

We start with the topology
\begin{equation} \label{twojet}
e^+ e^- \rightarrow jet + jet + \slashed{p}\,,
\end{equation}
where $\slashed{p}$ stands for missing energy and momentum in the
final state, i.e. the invariant mass of the two--jet system is
significantly smaller than the center--of--mass energy $\sqrt{s}$.
The extra Feynman diagrams contributing to this topology in our model
are shown in Fig.~\ref{fig:diag1}. As usual we neglect the Higgs
exchange diagrams since the $e^+e^-H$ coupling is tiny. Since the couplings,
$g_q$ and $g_\chi$, appear together in Fig.~\ref{fig:diag1}, the experiment
data bound the product of $g_q$ and $g_\chi$. Therefore, in the following
section, bounds on $\sqrt{g_qg_\chi}$ are shown.

\subsubsection{Analysis of LEP2 Data}
\setcounter{footnote}{0}

During the LEP2 period (data taken between 1995 and 2000, at center of
mass energy $161 \ {\rm GeV} \leq \sqrt{s} \leq 209 \ {\rm GeV}$)
ALEPH performed most searches for the topology (\ref{twojet}) in the
context of supersymmetric extensions of the SM. This includes searches
for the pair production of squarks \cite{barate1999searches,
  heister2003search} and neutralinos \cite{Barate:1997zx,
  Barate:1999fs, Barate:2000tu, abbiendi2000search,
  Heister:2003zk}. In addition, ALEPH searched for the production of
an invisibly decaying Higgs boson produced in association with an
on--shell $Z$ boson \cite{barate1999search, acciarri1998missing,
  Barate:1999uc}. Each of these searches uses dedicated cuts to
suppress the SM background.

We generally find that the data taken at higher energies have better
sensitivity to our model, {\em if} the event selection cuts are more
or less independent of $\sqrt{s}$. The cross section for the
four--body final state we are interested in depends quite sensitively
on the available phase space. Note also that the integrated luminosity
was higher at the higher energies. The total sensitivity is then
essentially determined by the data taken at higher energy. On the
contrary, if the cuts strongly depend on $\sqrt{s}$, the cut
efficiencies may vary strongly; in this case one should consider all
analyses together.

The neutralino searches fall in the second category.  The analyses of
the data taken at $\sqrt{s} = 161$ and $172$ GeV \cite{Barate:1997zx}
use quite different cuts than the analyses of the data taken at
$\sqrt{s} \geq 183$ GeV \cite{Barate:1999fs, abbiendi2000search,
  Barate:2000tu, Heister:2003zk}. At these higher energies, on--shell
production of two $Z$ bosons becomes possible. The high--energy
analyses impose a strong cut on the missing mass, which is designed to
remove the $Z\bar{\nu}\nu$ background. Unfortunately this cut by
itself excludes more than $90\%$ of our signal, leading to a total cut
efficiency of only about $2\%$. On the other hand, the lower energy
analyses use a cut on the visible mass, not on the missing
mass,\footnote{Note that in general there is no simple relation
  between the missing and the visible mass of a given event. The
  visible mass is defined as $M_{\rm vis}^2 = P_{\rm vis}^2$, where
  $P_{\rm vis}$ denotes the sum of the $4-$momenta of all ``visible''
  particles; only neutrinos and DSPs are counted as ``invisible''. The
  missing mass is defined by
  $M^2_{\rm miss} = ( P_{\rm init} - P_{\rm vis} )^2$, where
  $P_{\rm init}$ is the $4-$momentum of the initial state. In some
  kinematical configurations both the visible and the missing mass are
  small.} leading to a total cut efficiency of about $20\%$ for our
signal. The overall cross sections times luminosity at $\sqrt{s}= 161$
and $172$ GeV are, however, too small. We therefore find that the
analyses do not lead to significant bounds on our model.

For the invisibly decaying Higgs search, cut--based analyses were
published only for data with $\sqrt{s} \leq 183$ GeV
\cite{barate1999search}. There is a published search for this channel
using data taken at $\sqrt{s} = 189$ GeV \cite{Barate:1999uc}, but it
uses a Neural Network; since we cannot reproduce this analysis, we
cannot use it to constrain our model. ALEPH did not publish any search
for an invisibly decaying Higgs using data taken at $\sqrt{s} > 189$
GeV. When applied to our signal, the cuts used in the analyses
\cite{acciarri1998missing, barate1999search} at $\sqrt{s}$ between
$161$ and $183$ GeV have an efficiency of less than $10\%$. In this
case the most harmful cuts are those related to the thrust and the
reconstruction of the two jets. The relatively small cross sections,
low integrated luminosity and insufficient cut efficiencies again
imply that no meaningful constraints on our model can be derived.

We find the best sensitivity to our model when applying the cuts
optimized for searches for squark pair production. Here cut--based
analyses were published for the entire data set, including the highest
energies. The cuts have been listed in Sec.~7 of
ref.\cite{heister2003search}\footnote{The cuts for ``intermediate
  $\Delta M$'' usually turned out to give the tightest
  constraints. The influential cuts are $N_{\rm ch}>11$,
  $M_{\rm vis}>15$ GeV, $p_T/\sqrt{s}>4\%$,
  $E_{\rm vis}/\sqrt{s} < 70\%$, $E_{12}/\sqrt{s}<0.5\%$,
  $\cos\theta_{\rm miss}>0.8$, $\cos\theta_T>0.8$,
  $\Phi_{\rm acop}<176\degree$, $\Phi_{T}<177\degree$, 
  $E_{\rm Wedge}/\sqrt{s}<12.5\%$,
  ${\rm Thrust}<0.94$, $p_T/E_{\rm vis}>12.5\%$,
  $E_{\rm had}/\sqrt{s}<55\%$, $E_{\rm NH}/E_{\rm vis}<30\%$, and
  $E_{l1}^{30}/\sqrt{s}>1\%$. Here $N_{\rm ch}$ is the number of good
  tracks (i.e., of charged particles); $M_{\rm vis}$ is the invariant
  mass of the visible system, $E_{\rm vis}$ is its energy and $p_T$ is
  the absolute value of its transverse momentum, which is the same as
  the absolute value of the missing $p_T$; $\theta_{\rm miss}$ is
  the polar angle of the missing $p$ vector; $\theta_T$ is the polar
  angle of the thrust axis; $\Phi_{\rm acop}$ is the acoplanarity
  angle; $E_{\rm had}$ is the total measured energy excluding the
  contribution of identified charged leptons; and $E_{\rm Wedge}$ is the
  energy in a $30\degree$ azimuthal wedge around the missing transverse
  momentum. We also use many of these variables in the optimized cuts
  presented in Sec.~\ref{sec:4}, e.g. $E_{12}$, $E_{\rm NH}$,
  $E_{l1}^{30}$, $\Phi_T$, where their definition and physical significance are
  discussed.}; when applied to our model, they frequently lead to an
efficiency of $\geq 10\%$. This is still not ideal, but sufficient to
derive some meaningful constraints on the parameters of our model.

\begin{figure}[htb]
\includegraphics[width=0.5\textwidth]{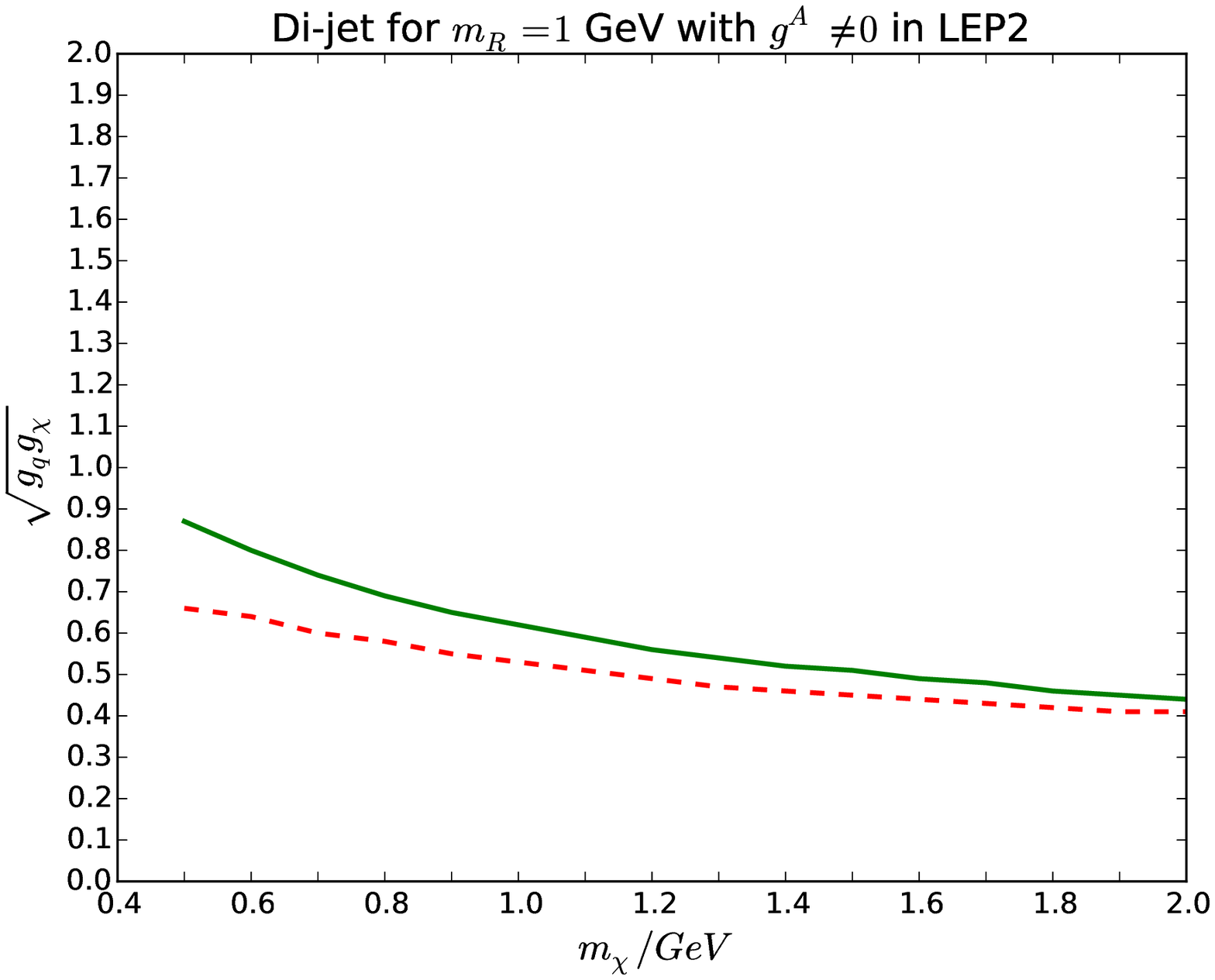}
\includegraphics[width=0.5\textwidth]{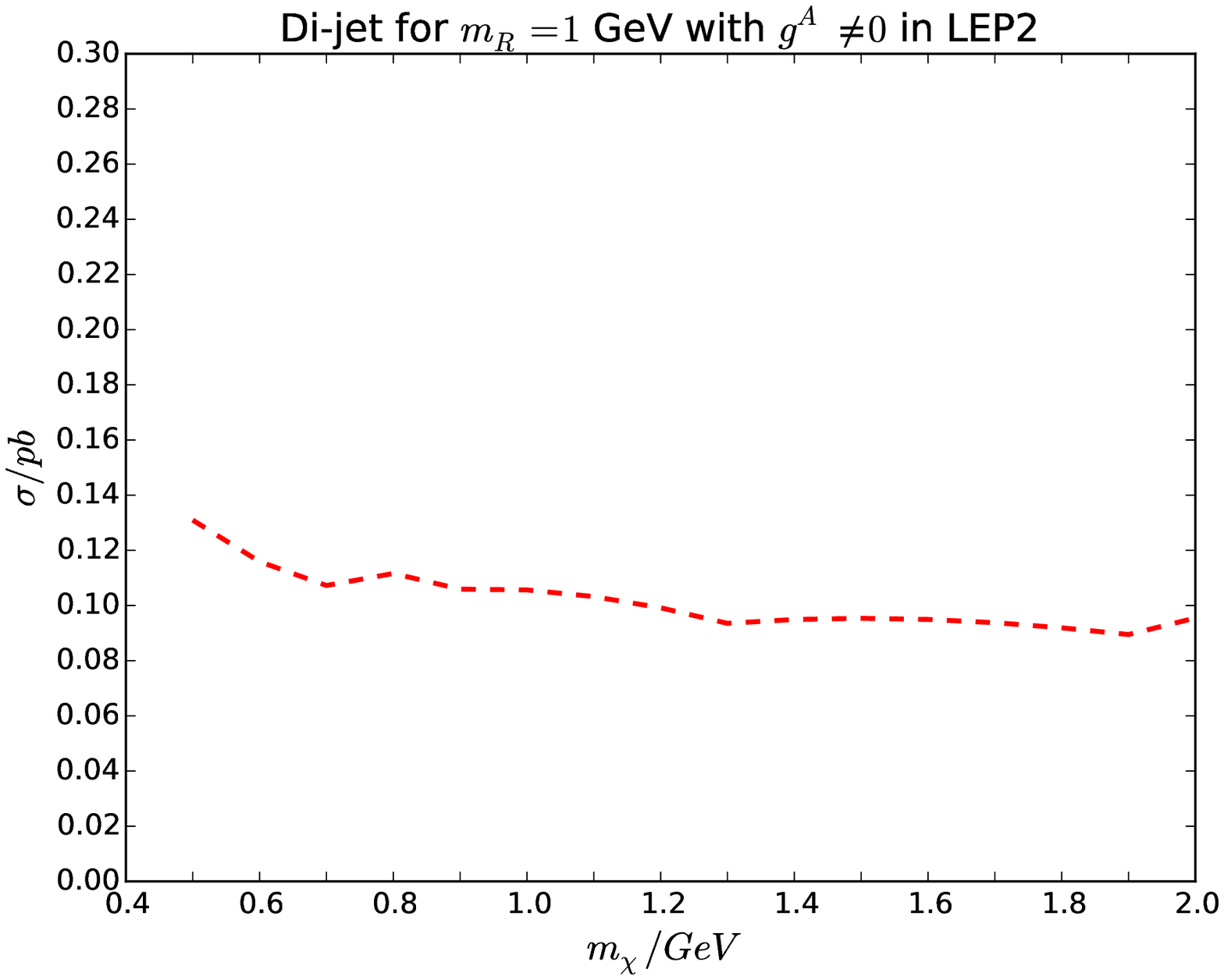}
\caption{The left frame depicts the bound on $\sqrt{g_qg_\chi}$. The solid
  (green) curve shows the bound on $g^A$ from the unitarity condition
  (\ref{unitarity}). The dashed (red) curve shows the bound on $g^A$ 
  from the combination of the unitarity condition and our recasting of
  the ALEPH squark pair search limits. The right frame shows the upper
  bound on the total signal cross section at $\sqrt{s} = 208$ GeV from our
  recasting of the ALEPH limits. In these figures the mass of the
  mediator $m_R = 1$ GeV. Here all vector couplings have been set
to zero, i.e. $g_q = g^A_q, \ g_\chi = g^A_\chi$.}
\label{fig:01}
\end{figure}

\begin{figure}[h!]
\includegraphics[width=0.5\textwidth]{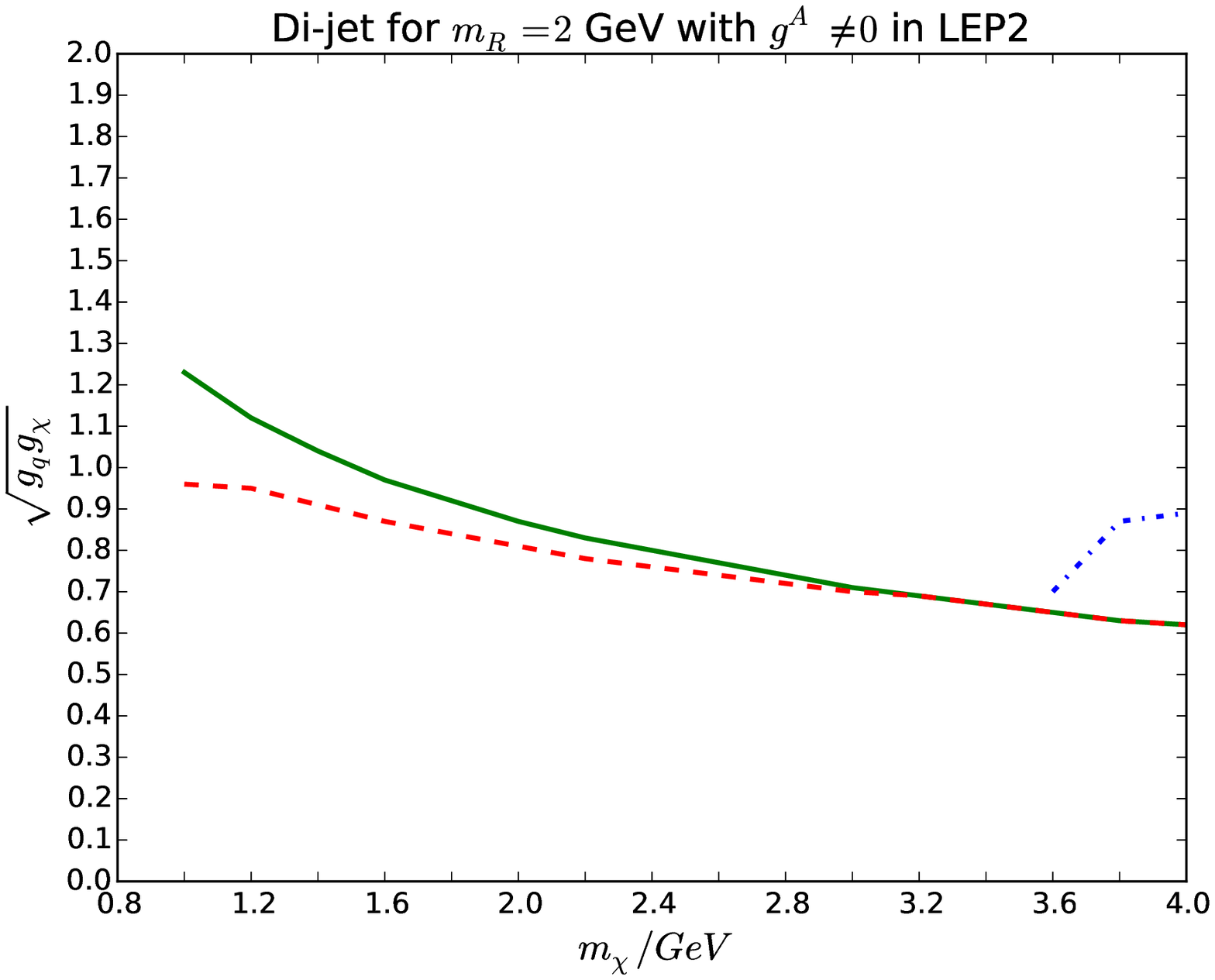}
\includegraphics[width=0.5\textwidth]{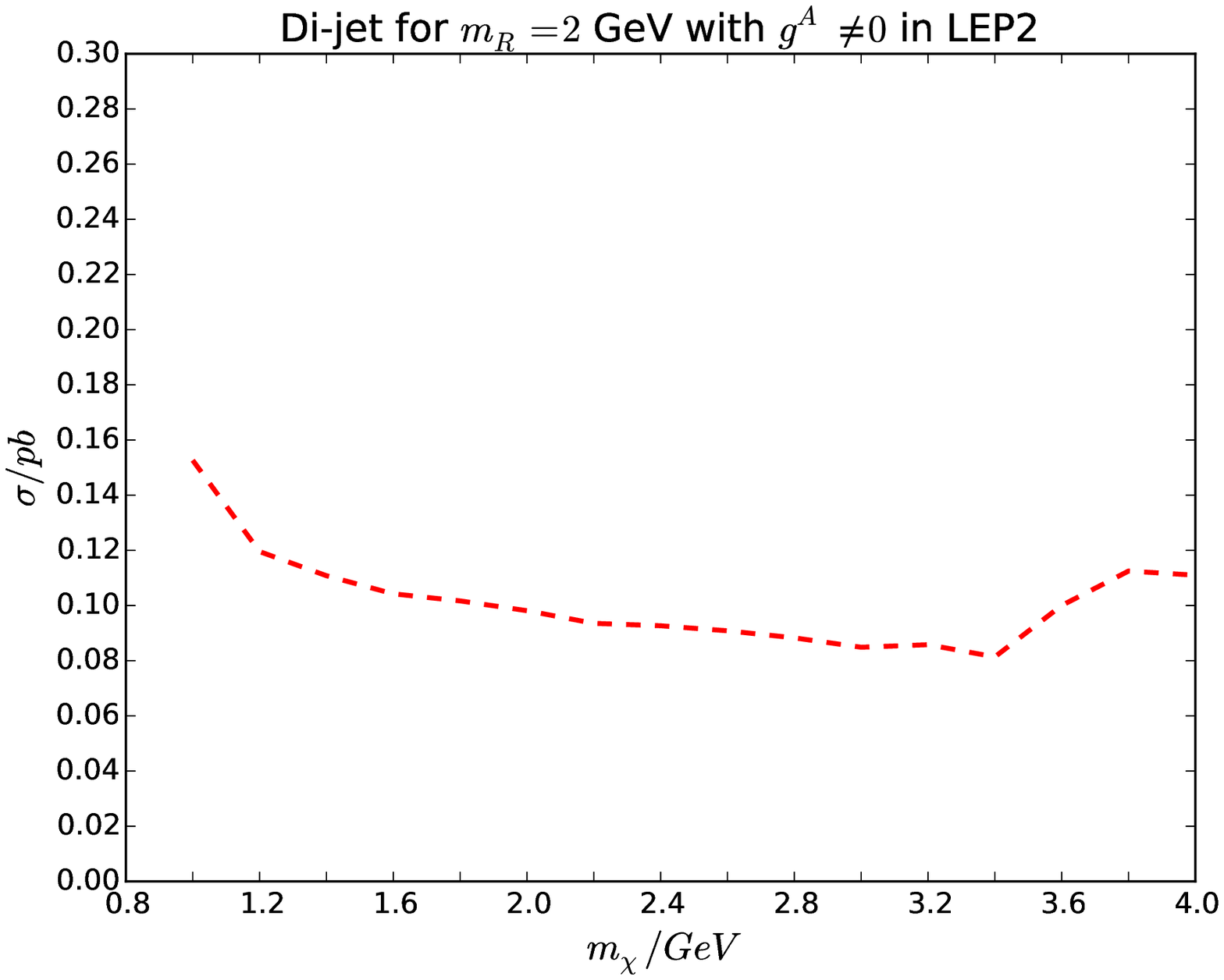} \\
\includegraphics[width=0.5\textwidth]{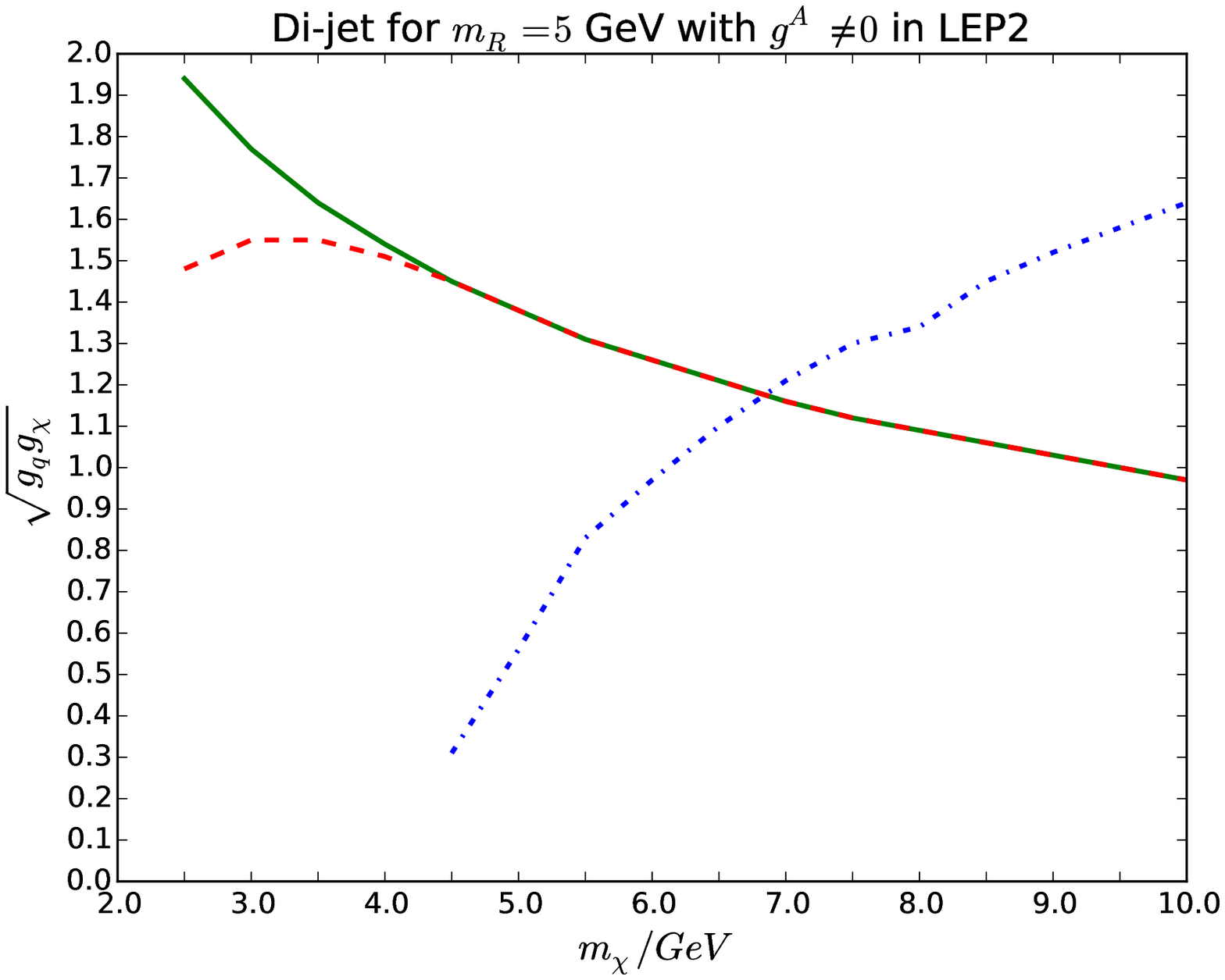}
\includegraphics[width=0.5\textwidth]{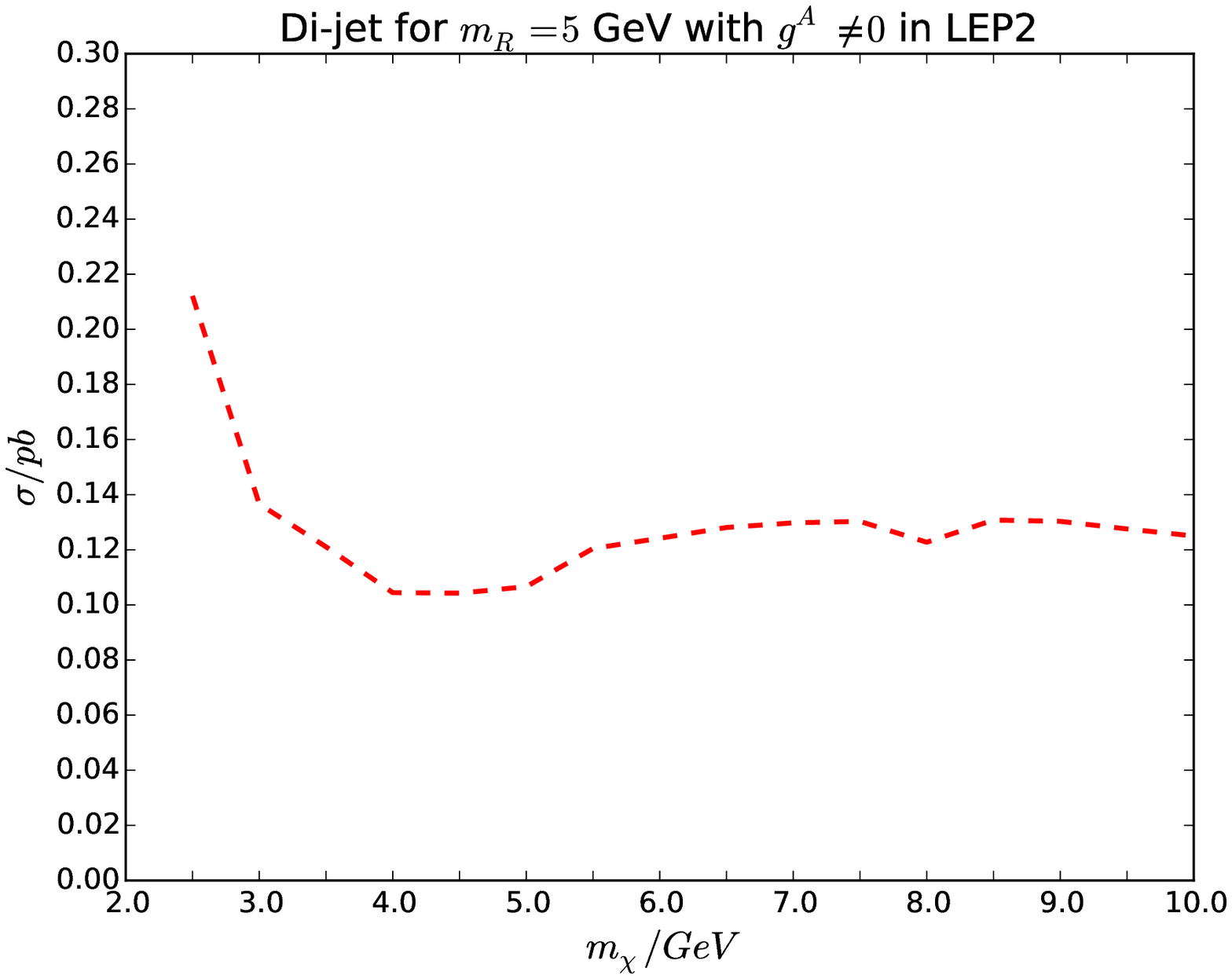}
\caption{The meaning of the curves in the right frames, and of the
  solid and dashed curves in the left frames, is as in
  Fig.~\ref{fig:01}, but for mediator mass $m_R = 2 \ (5)$ GeV in the
  top (bottom) frames. For $m_\chi \geq 3.6 \ (4.5)$ GeV the solid and
  dashed curves coincide, i.e. the unitarity condition gives the
  stronger bound on the axial vector coupling. The dotted (blue) lines
  show the upper bound on the vector coupling that we derive from the
  ALEPH search, i.e. for these curves, $g_q g_\chi = g^V_q g^V_\chi$;
  the axial vector couplings were chosen such that the unitarity limit
  is saturated. In this mass range the upper bound on the signal cross
  section shown in the right frames also uses the maximal axial vector
  coupling allowed by unitarity.}
\label{fig:03}
\end{figure}

\begin{figure}[htb]
  \includegraphics[width=0.5\textwidth]{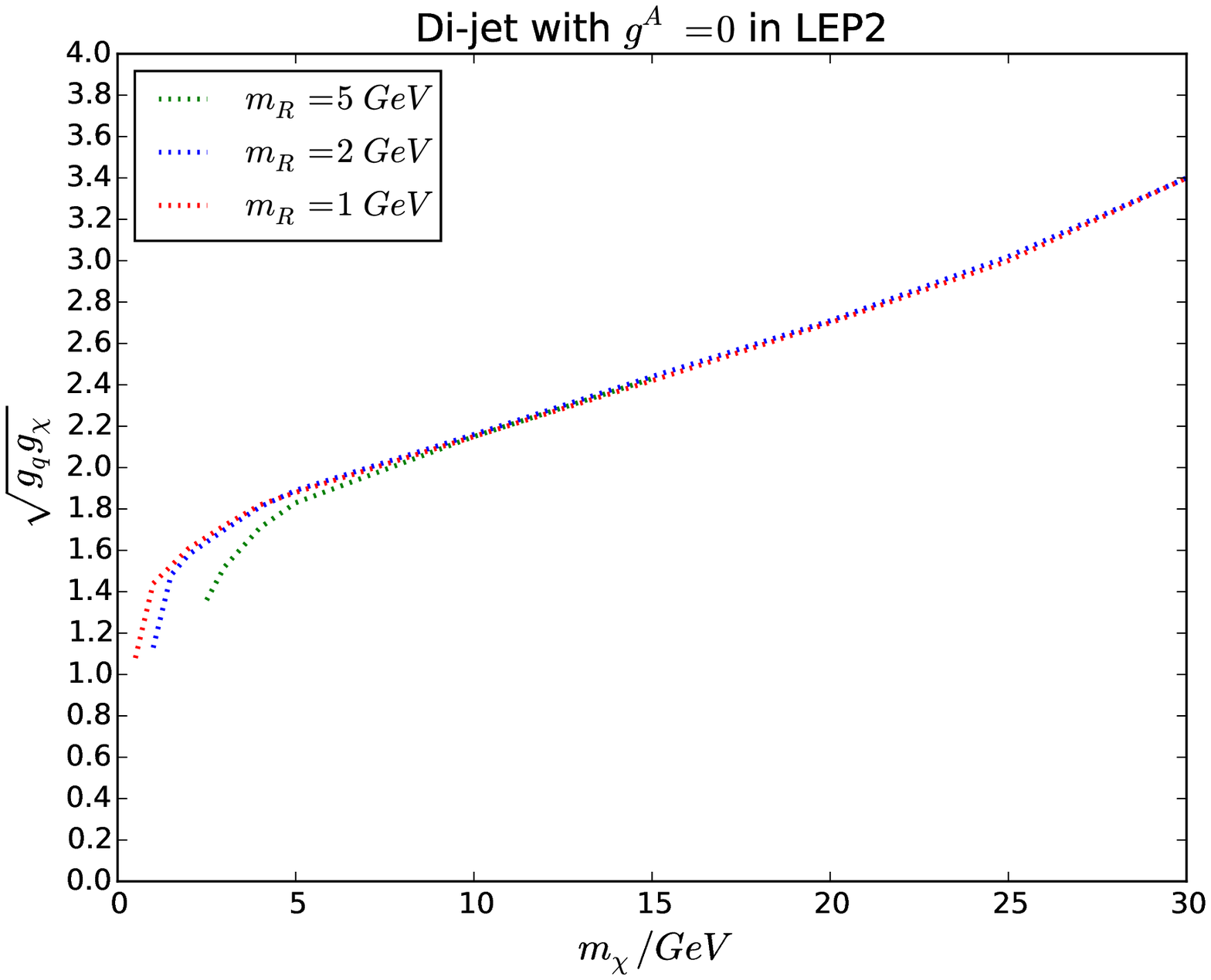}
  \includegraphics[width=0.5\textwidth]{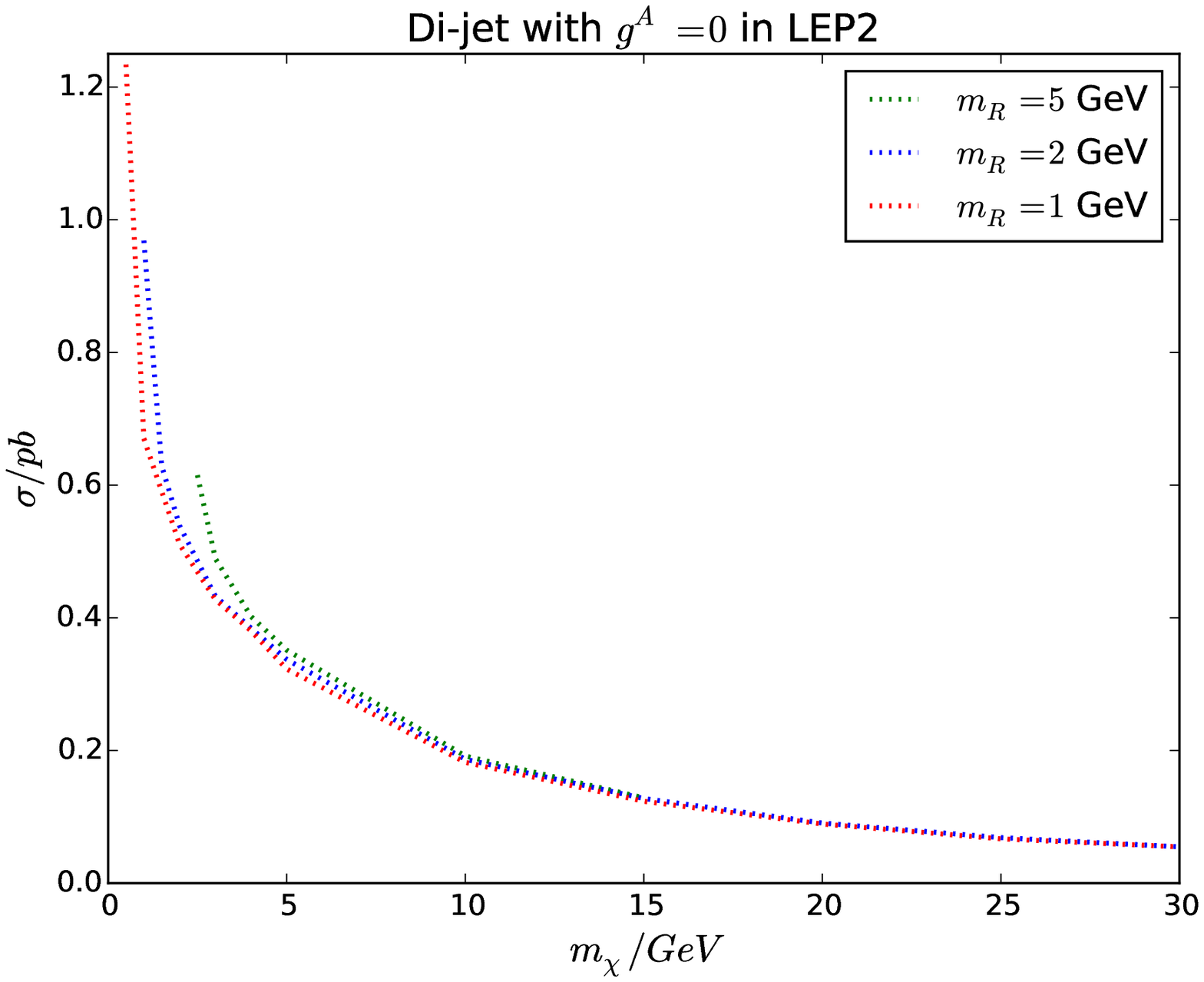}
  \caption{The left frame depicts the bound on $\sqrt{g_q^Vg_\chi^V}$, while the right
    one shows the upper bound on the total signal cross section at
    $\sqrt{s} = 208$ GeV from our recasting of the ALEPH limits; all
    axial vector couplings have been set to zero. The green, blue and
    red curves are for $m_R = 5\,,\ 2$ and $1$ GeV, respectively. For
    $m_\chi > 30$ GeV the bound on $g^V$ is weaker than the
    perturbativity condition (\ref{pert2}).}
  \label{fig:08}
\end{figure}

In this study we focus on the part of parameter space where on--shell
$R \rightarrow \chi \bar \chi$ decays are not allowed, i.e.
$m_\chi > m_R/2$, since otherwise ``monojet'' searches by the LHC
experiments \cite{Fox:2011pm,Khachatryan:2014rra,Chala:2015ama} give
much tighter constraints. In this part of parameter space our signal
process is a genuine $2 \rightarrow 4$ reaction, with rather low cross
section. We find that our recasting of the ALEPH squark searches does
not lead to significant constraints if $m_R \gsim 10$ GeV. In
Fig.~\ref{fig:01} and \ref{fig:03} we therefore show results for
$m_R = 1, \ 2$ and $5$ GeV, respectively, focusing on scenarios with
rather light DSP, $m_R/2 \leq m_\chi \leq 2 m_R$. We find that the
bounds on vector couplings are not as strong as those on the axial
vector couplings, and do not depend strongly on $m_R$. Therefore, for
nonzero $g^A$ and small $m_\chi$ we set the vector couplings to zero
and derive the upper bound on the axial vector coupling from the ALEPH
data in Figs.~\ref{fig:01} and \ref{fig:03}, while we show results for
$g^A=0$ separately in Fig.~\ref{fig:08}. For $m_\chi \gsim 4$ GeV and
$g^V = 0$ the resulting bound on $g^A$ is weaker than the unitarity
bound (\ref{unitarity}). In this case we set the axial vector coupling
such that the unitarity bound is saturated, and derive the resulting
upper limit on the vector coupling. This is the strongest possible
constraint on the vector coupling that can be derived from our
recasting of the ALEPH squark pair search. The larger sensitivity to
the axial vector coupling again comes from contributions
$\propto m_q m_\chi / m_R^2$ to the Feynman amplitude.

These terms dominate the cross section for $m_R = 1$ GeV
(Fig.~\ref{fig:01}). As a result, the bound on the coupling becomes
{\em stronger} as the DSP mass is increased. Evidently the enhanced
contribution from longitudinal $R$ exchange over--compensates the
reduction of the phase space. For the entire range of $m_\chi$ shown
the bound is stronger than the unitarity limit. Note that we show the
bounds on $\sqrt{g^A_\chi g^A_b}$, because the unitarity limit due to
$m_\chi$ is different compared to that due to $m_b$.  Moreover, all
vector couplings have been set to zero. Strictly speaking we would
have to allow some coupling at least to $s$ quarks in order to allow
$R$ to decay; however, vector couplings $\ll 1$ will not affect the
bound on the axial vector coupling. On the other hand, ${\cal O}(1)$
vector couplings would lead to a slightly stronger upper bound on the
axial vector coupling.

The results for $m_R = 2$ GeV (Fig.~\ref{fig:03}, top row) are
qualitatively rather similar, but the bound on the axial vector
coupling is weaker by a factor of about $1.5$. As a result, for
$m_\chi \geq 3.6$ GeV the upper bound on $g^A_\chi$ is actually set by
the unitarity constraint (\ref{unitarity}). At $m_\chi = 4$ GeV a
vector coupling as large as $0.89$ has been turned on in order to
saturate our recasting of the ALEPH bound, for axial vector coupling
at the unitarity limit. This leads to a slight increase of the upper
bound on the cross section, shown in the right frame, which otherwise
is very similar to the case with $m_R = 1$ GeV. Since in both cases
$m_R$ is much smaller than all other relevant energy scales in the
problem, in particular much smaller than the missing energy required
by the cuts, it is not surprising that the upper bound on the cross
section does not depend on $m_R$.

On the other hand, for $m_R = 5 \ {\rm GeV} \ \simeq m_b$
(Fig.~\ref{fig:03}, bottom row) the enhancement due to the exchange of
longitudinal $R-$bosons no longer suffices to over--compensate the
reduced phase space when $m_\chi$ is increased. The ALEPH squark
searches now permit quite large axial vector couplings even for
$m_\chi$ near $m_R/2$. For $m_\chi \geq 4.5$ GeV this bound again
becomes weaker than the unitarity constraint (\ref{unitarity}). Even
if we saturate this constraint, sizable vector couplings are allowed
by the ALEPH data, as shown by the dotted (blue) curve. Note that our
perturbativity bound (\ref{pert2}) requires $g^V_q \leq 2.5$ for
$m_R = 5$ GeV. Fig.~\ref{fig:03} shows that our recasting of the ALEPH
squark search limits leads to stronger upper bounds on this coupling if
$m_\chi \leq 10$ GeV.

Having considered nonzero $g^A$, the bounds on $g^V$ for vanishing
$g^A$ are shown in Fig.~\ref{fig:08}. Evidently the constraints on
$g^V$ are much weaker than those on $g^A$. Recall, however, that $g^V$
is not constrained by the unitarity condition. The upper bound on
$g^V$ is therefore set by LEP2 data for $m_\chi \lsim 30$ GeV; at even
larger DSP masses, the LEP2 bound becomes weaker than the
perturbativity condition (\ref{pert2}). Another noticeable property is
that for $m_\chi > 10$ GeV the upper bound on the vector coupling is
nearly the same for our three choices of $m_R$, as is the bound
on total cross section. This is due to the fact that the transverse $R$
propagator becomes independent of $m_R$ once $(2 m_\chi)^2 \gg m_R^2$.

\begin{figure}[htb]
\includegraphics[width=0.5\textwidth]{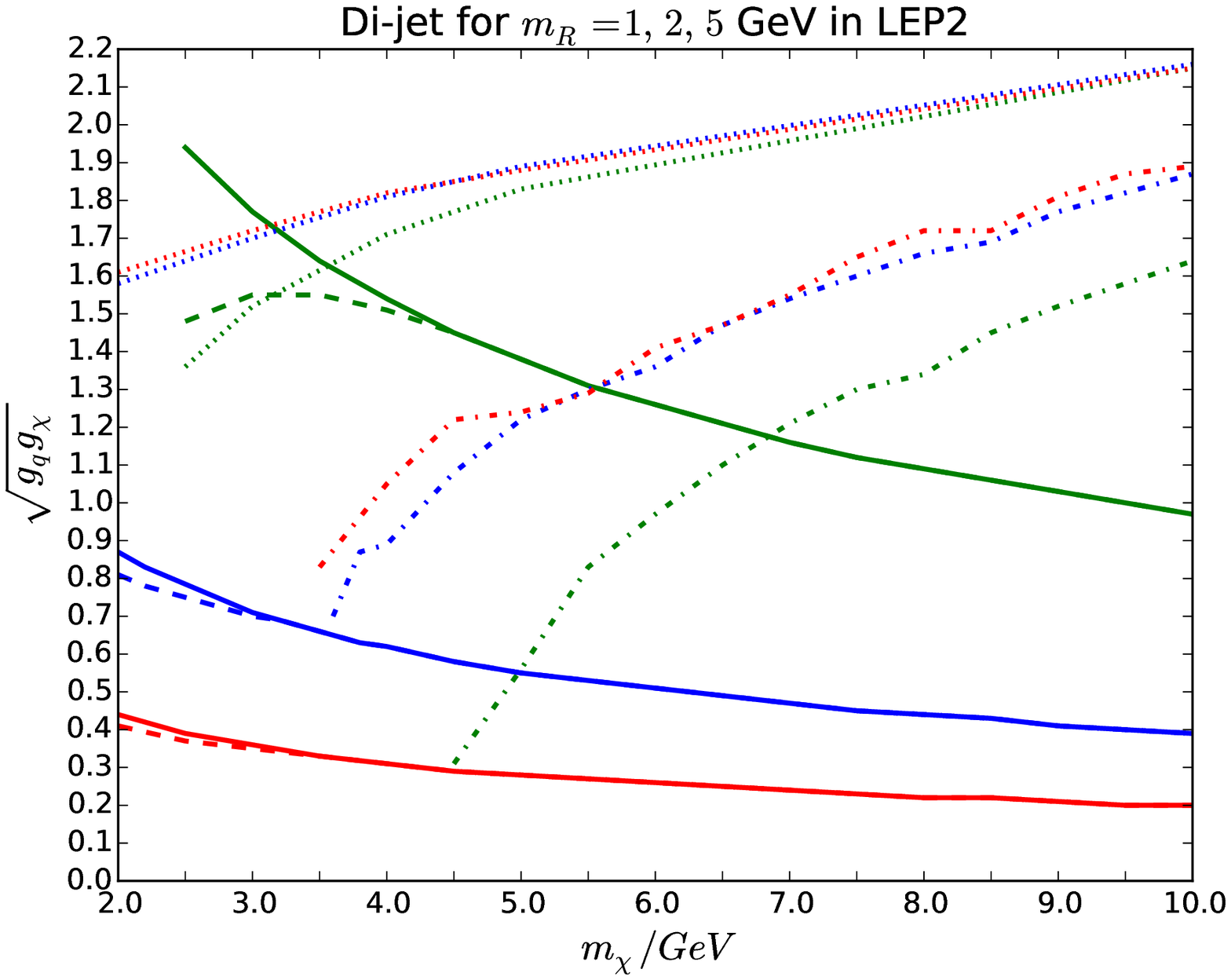}
\includegraphics[width=0.5\textwidth]{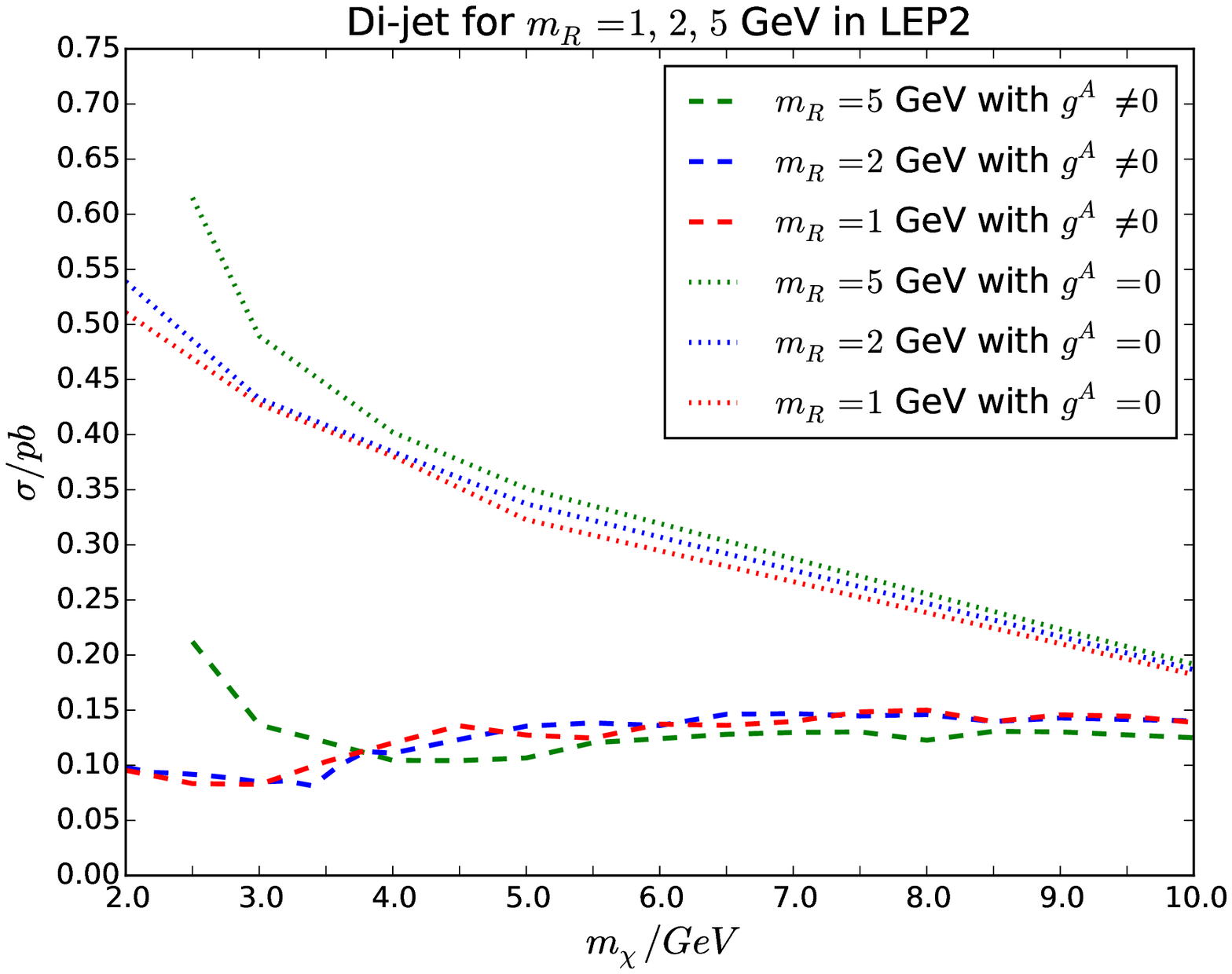}
\caption{The left frame shows upper bounds on the couplings of the
  mediator. The solid lines are from the unitarity condition
  (\ref{unitarity}), while the dashed lines are from our recasting of
  the ALEPH squark search; both sets of curves refer to axial vector
  couplings.  If the unitarity constraint is stronger, we allow
  non--vanishing vector couplings; their upper bounds, derived from
  the ALEPH data, are shown by the dot--dashed curves. If $g^A=0$, the
  unitarity condition are irrelevant, and the bounds on $g^V$ become
  weaker. They are shown by the dotted curves. The green, blue and red
  curves are for $m_R = 5\,,\ 2$ and $1$ GeV, respectively. The right
  frame shows the corresponding upper bound on the total signal cross
  section before cuts at $\sqrt{s} = 208$ GeV.}
\label{fig:04}
\end{figure}

In Figs.~\ref{fig:01} and \ref{fig:03} we extended $m_\chi$ only to
$m_\chi = 2 m_R$. In Fig.~\ref{fig:04} we show upper bounds on the
couplings (left) and on the total cross section (right) for the same
values of $m_R$, but for $m_\chi$ between $2$ and $10$ GeV. Moreover,
we also compare the bounds on $g^V$ for $g^A=0$ (dotted curves) to the
bounds on $g^V$ with $g^A$ chosen to saturate its upper bound (dashed
curves), which is set by the unitarity condition (\ref{unitarity})
once $m_\chi > 4$ GeV. In this case
$g^A_\chi g^A_b \propto m_R^2 / (m_\chi m_b)$, so that the
contribution from longitudinal $R$ exchange becomes largely
independent of both $m_\chi$ and $m_R$ once $(2 m_\chi)^2 \gg m_R^2$.
Note that the axial vector coupling also contributes to the exchange
of transverse $R$ bosons. This contribution simply scales like
$(g^A)^2$, and is thus significant only for $m_R = 5$ GeV where
unitarity allows relatively large axial vector couplings. This
explains why the upper bound on $g^V$ with maximal $g^A$ is stronger
for $m_R = 5$ GeV than for the smaller values of $m_R$. In contrast,
if $g^A = 0$ the bound on $g^V$ becomes independent of $m_R$ once
$(2m_\chi)^2 \gg m_R^2$, as we saw above. Overall Fig.~\ref{fig:04}
shows that the effect of $g^A$ can be significant even if it is much
smaller than $g^V$.

The right frame of Fig.~\ref{fig:04} again shows that the upper bound
on the cross section becomes independent of $m_R$ once
$(2 m_\chi)^2 \gg m_R^2$. We also see that for light $R$ and
$g^A \neq 0$, the upper bound on the cross section increases by nearly
a factor of two once $m_\chi > 4$ GeV; evidently the cut efficiency
becomes smaller. This coincides with the range of $\chi$ masses where
the bound on the axial vector coupling is set by the unitarity
constraint, so that the limit we derive from the ALEPH data can only
be saturated by also including sizable vector couplings. The main
observation is that the cut efficiency is much smaller if the process
proceeds dominantly by vector coupling. For example, for
$m_R = m_\chi = 1$ GeV, we find cut efficiencies between $1.5$ and
$3.5$\% for pure vector coupling, with couplings to $b$ quarks
yielding the highest sensitivity. In contrast, if the cross section is
dominated by the axial vector coupling to $b$ quarks the efficiency
increases to $15$\%. This is at least partly due to the fact that the
$\chi \bar \chi$ pair has to be in a $P-$ wave in the $R$ rest frame
if the $\chi \bar \chi R$ coupling is purely axial vector, whereas a
vector coupling allows $S-$wave contributions. The $P-$wave has a
larger $\chi \bar \chi$ invariant mass, making it easier to pass cuts
related to the missing mass or missing energy. The cut efficiency
increases with increasing $m_\chi$, which of course also implies
larger $\chi \bar\chi$ invariant mass. However, even here pure vector
couplings lead to lower cut efficiency.  For example, for
$m_\chi = 10$ GeV, i.e. at the end of the range shown in
Fig.~\ref{fig:04}, we find a cut efficiency of just under $10$\% if
$g^A$ saturates the unitarity bound, with little dependence on $m_R$;
if $g^A = 0$, the cut efficiency is only about $7.5$\%.

We also find reduced cut efficiency if $m_\chi$ is only slightly above
$m_R/2$. In this case configurations where the $R$ boson is only
slightly off--shell, i.e. configurations with small $\chi \bar\chi$
invariant mass, are even more strongly preferred dynamically than for
larger values of the ratio $m_\chi / m_R$. This again leads to a
reduced efficiency for cuts related to the missing mass.

\subsubsection{Analysis of LEP1 Data}
\setcounter{footnote}{0}

Searches for the final state (\ref{twojet}) were also performed at
LEP1, the first period of operating the LEP collider (1989 to 1994),
with $\sqrt{s} \simeq 91 \ {\rm GeV} \ \simeq m_Z$ \cite{Decamp:1991uy,
  Buskulic:1996hz}. These analyses searched for $H \nu \bar \nu$
production where $H$ is the SM Higgs boson which is assumed to decay
hadronically; this final state yielded the strongest lower bound on
$m_H$ that could be derived from a single LEP1 analysis.

Since the exchanged $Z$ boson is now nearly on--shell, for not too
large values of $m_\chi$ the total signal cross section is much larger
than at LEP2. Moreover, the physics background at
$\sqrt{s} \simeq m_Z$ is much smaller than at $\sqrt{s} \simeq 200$
GeV. In particular, the $W^+W^-$ and $ZZ$ backgrounds did not exist at
LEP1. Therefore, less severe cuts were needed at LEP1, so the cut
efficiency of our signal can be expected to be higher than for the
LEP2 analyses. These two effects over--compensate the about three
times smaller total luminosity accumulated at LEP1. At least for not
too large DSP mass we therefore expect LEP1 data to lead to stronger
constraints on the couplings of our model than LEP2 data.

In \cite{Decamp:1991uy} the cuts and the number of selected events are
not given in detail. We therefore cannot recast this
analysis. Fortunately it is superseded by \cite{Buskulic:1996hz},
where all applied cuts and the number of selected events are
listed\footnote{The influential cuts are: $N_{\rm ch} > 7$,
  $M_{\rm vis}<70$ GeV, 
  $p_{\rm CH}/\sqrt{s}>0.1$, $E_{30}/E_{\rm vis} > 60\%$, $E_{12}<3$
  GeV, $\theta_{\rm acol} < 165\degree$, $M_{\rm vis} > 25$ GeV when
  $p_T/\sqrt{s}<10\%$, $M_{\rm thrust}^{1,2}>2.5$ GeV,
  $\sum_{3j}\theta_{jj} < 342\degree$, $\Phi_{\rm acop} < 159\degree$,
  and $\Theta_{\rm miss}^{\rm iso} > 31\degree$. Here $p_{\rm CH}$ is
  the scalar sum of the charged particle momenta; $E_{30}$ is the
  energy measured at more than $30\degree$ from the beam axis;
  $M_{\rm thrust}^{1,2}$ is the invariant masses measured in both
  hemispheres according to the plane perpendicular to the thrust axis;
  and $\Theta_{\rm miss}^{\rm iso}$ is the largest cone around missing
  momentum vector containing energy less than $1$ GeV. The other
  variables have already been defined in the LEP2 analysis described
  in Sec.~3.1.1.}. Unfortunately there is some uncertainty regarding
the precise jet definition that has been used. One of the cuts
requires to reconstruct the final state as exactly three jets. We
found that the results differ slightly for different jet algorithms.
Moreover, occasionally the reconstruction of the event as three--jet
event does not work; we discard such events. However, both the effect
of having to discard events that cannot be described as three--jet
events, and the differences between final results using different jet
algorithms, are quite small, probably smaller than the effects of
ignoring detector smearing, as we do. In the results presented below
we use the $k_T$ based Durham algorithm, which was the algorithm of
choice for LEP2 analyses.

\begin{figure}[thb]
  \includegraphics[width=0.5\textwidth]{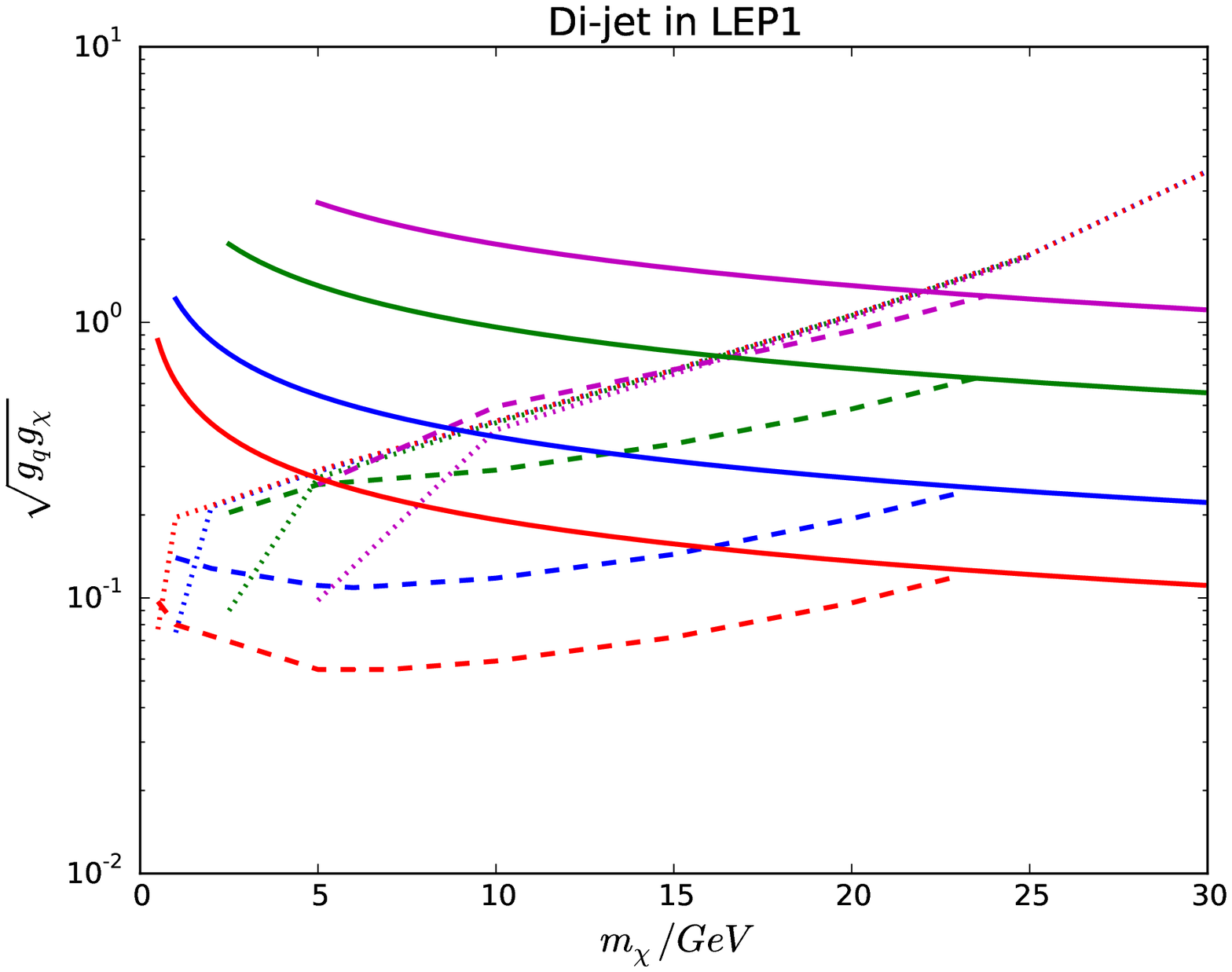}
  \includegraphics[width=0.5\textwidth]{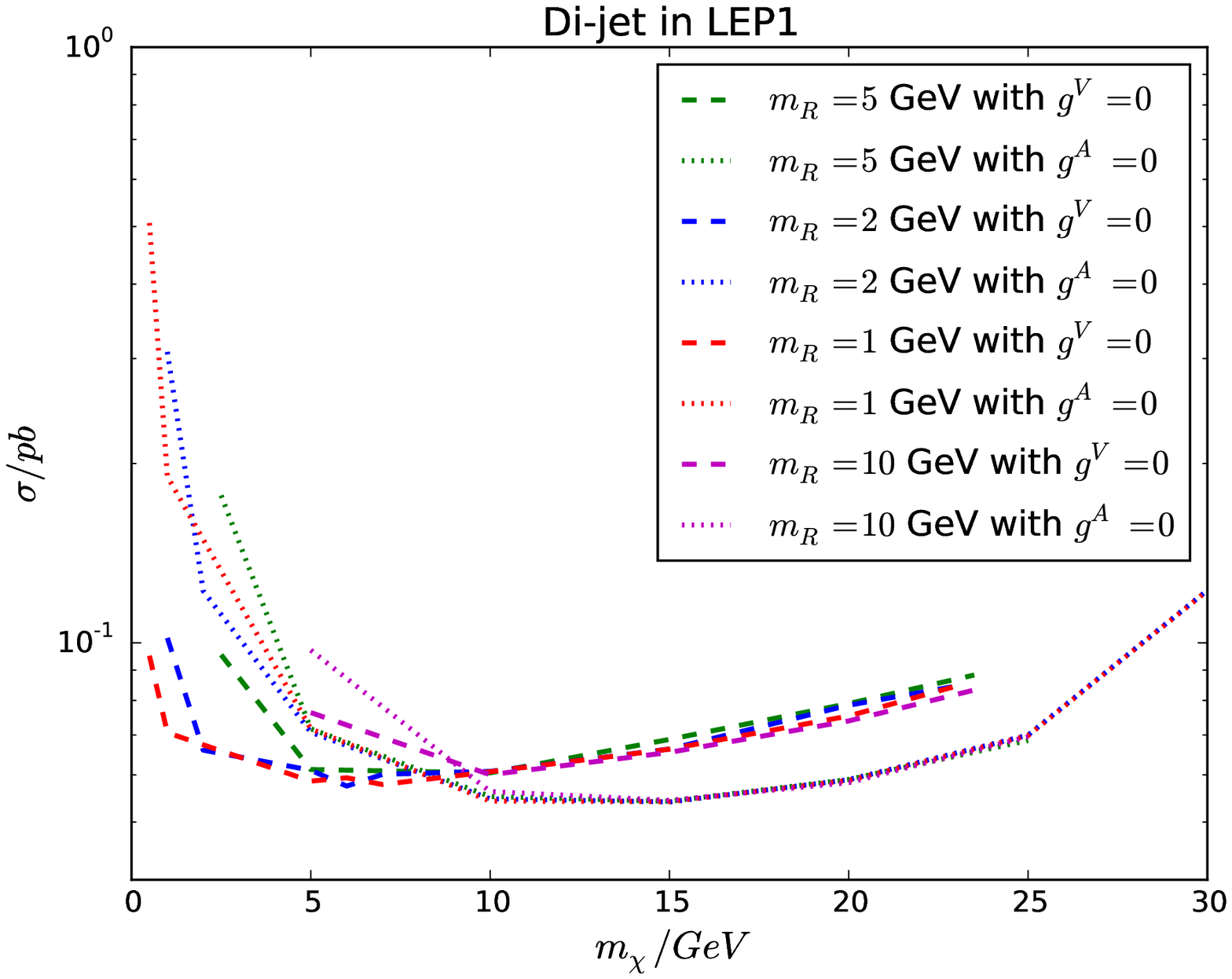}
  \caption{The left frame depicts bounds on couplings of the
    mediator. The solid lines show the unitarity bound on the axial
    vector coupling. The other curves depict bounds from our recasting
    of the ALEPH LEP1 limits. The dotted lines are the upper limits on
    $g^V$ for $g^A = 0$, while the dashed lines are the upper limits
    on $g^A$ for $g^V = 0$. The purple, green, blue and red curves are
    for $m_R = 10\,,\ 5\,,\ 2$ and $1$ GeV, respectively. For
    $m_\chi > 30$ GeV the bound on $g^V$ is always weaker than the
    perturbativity condition (\ref{pert2}). For $m_\chi > 23$ GeV the
    LEP1 bound on $g^A$ is weaker than the unitarity condition
    (\ref{unitarity}), and is therefore not shown any more. The right
    frame shows the upper bound on the total signal cross section at
    $\sqrt{s} = 91$ GeV; we use the same conventions as in the left
    frame.}
  \label{fig:09}
\end{figure}

The results are shown in Fig.~\ref{fig:09}. Evidently for
$m_\chi < 20$ GeV the bounds from LEP1 data are much better than those
from the published analyses of LEP2 data. For larger DSP mass,
however, the phase space constraints become quite severe at LEP1
energy, and hence these data quickly lose sensitivity. For pure axial
vector coupling the upper bound on the coupling we derive from our
recasting of the LEP1 data saturates the unitarity constraint at
$m_\chi \simeq 23$ GeV, with larger $m_R$ yielding a slightly larger
range of $m_\chi$ where the experimental bound is below the unitarity
limit. This can be understood as follows. The larger $m_R$, the larger
the axial vector coupling allowed by unitarity.  The contribution from
longitudinal $R$ exchange is again independent of $m_R$ if the
unitarity limit is saturated, but the contribution from the exchange
of transversely polarized $R$ bosons increases with increasing $g^A$,
and hence becomes significant only for larger $m_R$. Note also that
for small $m_R$ the bound on $g^A$ at first becomes stronger as
$m_\chi$ is increased from its minimal value, which we took to be just
above $m_R/2$ as before. As in Figs.~\ref{fig:01} and \ref{fig:03}
this is due to the exchange of longitudinal $R$ bosons.

In contrast, for $g^A = 0$ the bounds on $g^V$ are strongest for the
smallest value of $m_\chi$, where the $R$ boson only needs to be
slightly off--shell. The steep rise of the dotted curves towards small
$m_\chi$ in the right frame shows that the cut efficiency decreases,
but this is over--compensated by the increase of the total cross
section. Once $m_\chi > m_R$, the bound on $g^V$ again becomes
largely independent of $m_R$, and is (coincidentally) quite close to
the bound on $g^A$ for $m_R = 10$ GeV. The ``experimental'' upper
bound on $g^V$ becomes worse than the perturbativity constraint
(\ref{pert2}) once $m_\chi > 30$ GeV.

If $g^A \neq 0$ and $g^V = 0$ the cut efficiency of our signal is
generally higher than $20\%$. For
$5\ {\rm GeV} < m_\chi < 10 \ {\rm GeV}$ the cut efficiency is even
higher than $30\%$, and reaches the highest point of $32\%$ for
$m_\chi$ between $6$ GeV and $7$ GeV. For $g^V \neq 0, g^A = 0$ and
relatively small $m_\chi \gsim m_R/2$ the cut efficiency is again
less, typically around $10\%$, which is similar to the efficiency for
the LEP2 squark pair search. However, it quickly increases for larger
$m_\chi$, reaching $35\%$ for $m_\chi\simeq 15$ GeV. Moreover, for
$m_\chi > 10$ GeV the cut efficiency is now actually higher for pure
vector coupling than for pure axial vector coupling. This is opposite
to the results shown in Fig.~\ref{fig:04} for LEP2 energies. The LEP1
analysis mostly employs cuts on angular variables, and does not
contain any explicit cut on the invisible mass or energy; recall that
such cuts play a prominent role in the corresponding analysis of LEP2
data.

\subsection{Four Jet Analysis}
\setcounter{footnote}{0}

We now turn to a discussion of the $4-$jet final state. The signal
again comes from the diagrams shown in Fig.~\ref{fig:diag1}, except
that the (real or virtual) $R-$boson now decays into a $q \bar q$ pair
rather than a $\chi \bar\chi$ pair. As a result, at tree--level the
cross section now only depends on the couplings of $R$ to quarks. We
compute the signal by squaring the $R-$exchange contribution, i.e. we
neglect interference between $R-$exchange and SM contributions. Note
that the interference with the dominant (gluon exchange) SM
contribution to the four quark final state is color
suppressed\footnote{Denote the final state by
  $q(k_1) \bar q(k_2) q'(k_3) \bar q'(k_4)$, where $q'$ may be a
  different flavor from $q$. The gluon exchange contribution where
  $q'(k_3) \bar q'(k_4)$ results from the splitting of a virtual gluon
  then only interferes with the $R$ exchange contributions where
  $q'(k_3) \bar q(k_2)$ or $q(k_1) \bar q'(k_4)$ originate from the
  decay of the $R$ boson. Evidently this is possible only if $q' = q$,
  i.e. for final states with two identical $q \bar q$ pairs. Moreover,
  the interference gets a color factor of $1$, compared to a factor
  $N_c^2 = 9$ for the squared $R$ exchange diagram. We checked
  explicitly for some combinations of parameters that the interference
  terms change the total cross section only by a few percent.};
moreover, the total SM contribution to four parton final states is
dominated by $q \bar q g g$ production, where $g$ stands for a gluon.

There are several ALEPH analyses involving $4-$jet final states.  Some
are optimized to detect $W^+W^-$ or $ZZ$ final states. These are part
of the background for us; hence these analyses cannot be used to
derive useful bounds on the couplings of our model. The earliest ALEPH
analyses of the $4-$jet final state in the LEP2 era had very low
luminosity \cite{Buskulic:1996hx} or did not veto $ZZ$ events
\cite{Barate:1997tz}, and are hence also only of limited usefulness
for our purpose.

In contrast, the searches for neutral Higgs bosons, either in pairs or
in association with a $Z$ boson, investigate final states that are at
least somewhat similar to ours. More importantly, they include cuts
that attempt to minimize non--Higgs SM backgrounds, both from
electroweak and from QCD sources. The related analyses cover the
entire LEP2 energy range, from $\sqrt{s} = 133$ to $209$ GeV
\cite{Barate:1997rh, Barate:1997mb, Barate:1998gw, Barate:2000na,
  Nielsen:1999ct, Barate:2000zr, Barate:2000ts,Heister:2001kr}. The
analyses of the data taken at $\sqrt{s} \leq 172$ GeV all use similar
cuts, while the analyses of data taken at $\sqrt{s} \geq 183$ GeV
apply another group of cuts in order to reduce $W^+W^-$ and $ZZ$
backgrounds. The first group of analyses turns out to be essentially
useless for us, due to the rather low energy and comparatively small
integrated Luminosity.

\begin{figure}[htb]
\includegraphics[width=0.5\textwidth]{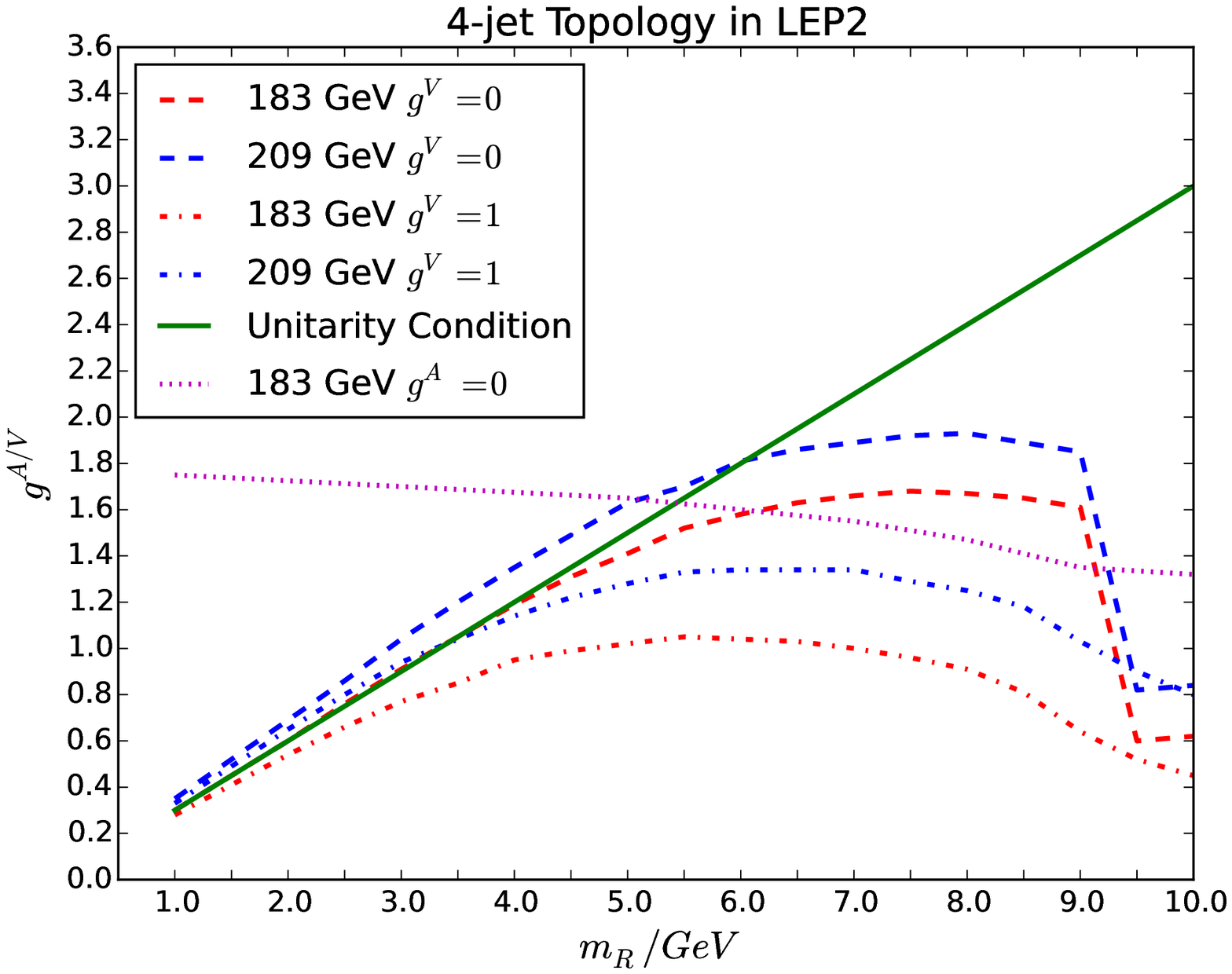}
\includegraphics[width=0.5\textwidth]{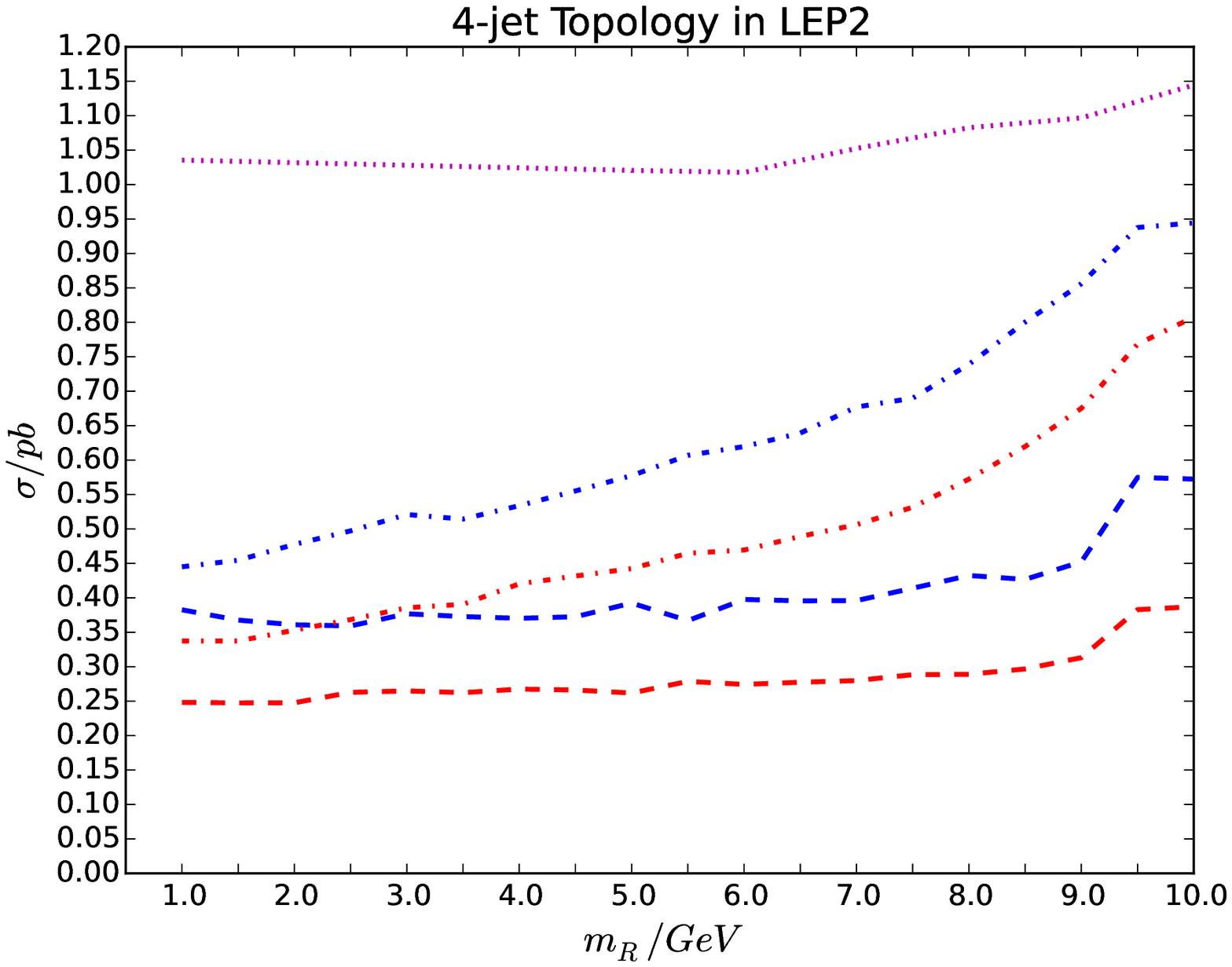}
\caption{Bounds on the (axial) vector coupling to $b$ quarks (left)
  and total cross section (right) we derive from our recasting of the
  ALEPH Higgs searches in the $4-$jet channel. In both frames the
  lower (red) curves correspond to data taken at $\sqrt{s} = 183$ GeV,
  while the upper (blue) curves correspond to data taken at
  $\sqrt{s} = 209$ GeV. The dashed curves have been obtained with
  vanishing vector couplings, while the dot--dashed curves are for
  $g^V_q = 1$. The dotted magenta curves show the upper bounds for
  vanishing axial vector couplings. The solid (green) curve in the
  left frame shows the upper bound on $g^A_b$ from the unitarity
  constraint (\ref{unitarity}) applied to the $b$ quark.}
\label{fig:05}
\end{figure}

However, the data taken at $\sqrt{s} \geq 183$ GeV do allow to impose
meaningful constraints on our model. Although the cuts applied in
these analyses are similar, the slight changes still influence the
final efficiencies. We find the highest efficiency, of about $27\%$
with little dependence on $m_R$, for the cuts applied to the data
taken at $\sqrt{s}=183$ GeV \cite{Barate:1998gw}\footnote{The
  influential cuts are: at least $2$ $b-$jets, $N_{\rm ch} > 7$,
  $\min(\cos\theta_{ij} + \cos\theta_{kl}) < -1.3$ ($ijkl$ label the
  four jets), $\min(\sum_{i=1}^4\theta_{jj}^i) > 350\degree$,
  and either $y_{34}>(2.9-\#b-{\rm jets})/9.5$
  (transition from 4 to 3 jets through Durham algorithm), or
  $m_{12}>78$ GeV, $m_{34}>55$ GeV, and $y_{34}>0.008$. Here
  $\theta_{ij}$ is the opening angle between jets $i$ and $j$,
  $m_{ij}$ is the invariant mass of the system of jets $i$ and $j$;
  $\theta^i_{jj}$ is any one of these six opening angles, with the sum
  going over the smallest four; $\#b-{\rm jets}$ is the number of
  tagged $b-$jets.}, where the $Z$ pair background is still very
small. At the highest energy the efficiency falls to about $21$ to
$22\%$. As a result, the strongest bound can be derived from the ALEPH
analysis of the data taken at $\sqrt{s} = 183$ GeV. This is shown in
Fig.~\ref{fig:05}.

The 183 GeV analysis performs quite well. For pure axial vector
coupling (dashed curves) the final cut efficiency for our signal is
actually as good as the one for the all--hadronic $ZH$ signal for
which this analysis was originally designed. This leads to quite
stringent bounds, in particular for small $m_R$, where it is
significantly stronger than that from the $2-$jet plus missing energy
analysis of LEP2 data described in the previous Section even for small
$m_\chi$, if we assume $g_q=g_\chi$; of course, the constraints we
derive from the analysis of the four jet final state are independent
of $m_\chi$ and $g_\chi$, as long as $m_\chi > m_R/2$. However, for
vanishing vector couplings our ``experimental'' bound on $g^A_b$ is
still slightly weaker than the one derived from the unitarity
constraint (\ref{unitarity}) applied to the $b$ quark, where we used
$m_b(m_b) = 4.25$ GeV.

Turning on a vector coupling $g^V_q = 1$ for $q=s,c,b$ reduces the cut
efficiency somewhat; this leads to increased upper bounds on the total
cross section. This is presumably again due to the $P-$wave nature of
the $q \bar q$ pair that originates from the ``decay'' of the virtual
$R$ boson via an axial vector coupling, which leads to a larger
separation between these two partons, and hence better separated
jets. Nevertheless the resulting upper bound on $g^A_b$ that we derive
from the $183$ GeV analysis is now better than the one from the
unitarity condition. This is in particular true for larger $m_R$; the
vector contribution depends less strongly on the mass of the mediator,
since there are no terms $\propto m_b^2/m_R^2$ in this case. For small
$m_R$ the upper bound on a pure vector coupling is rather weak, but
still stronger than the perturbativity limit (\ref{pert2}).

\begin{figure}[htb]
\includegraphics[width=0.5\textwidth]{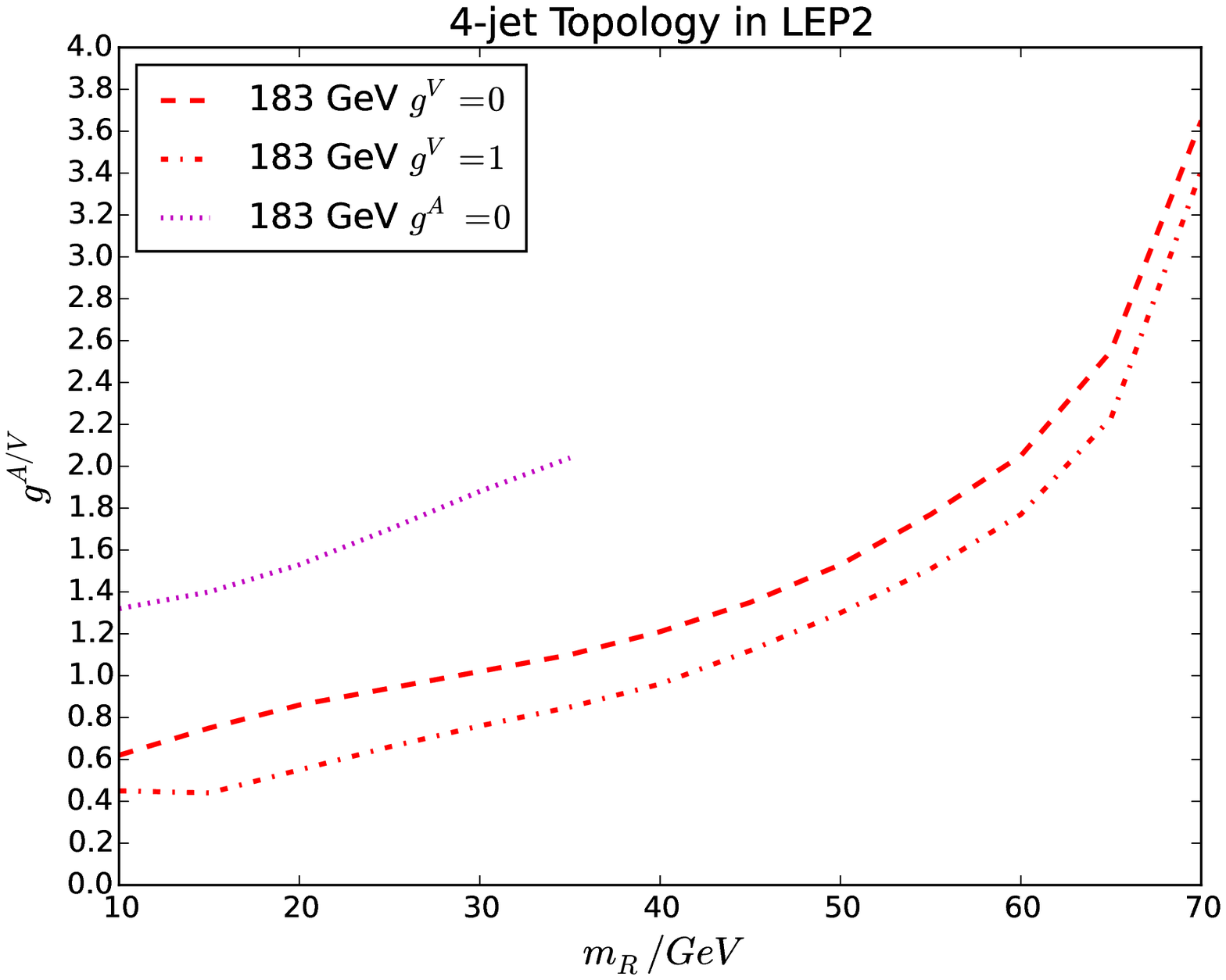}
\includegraphics[width=0.5\textwidth]{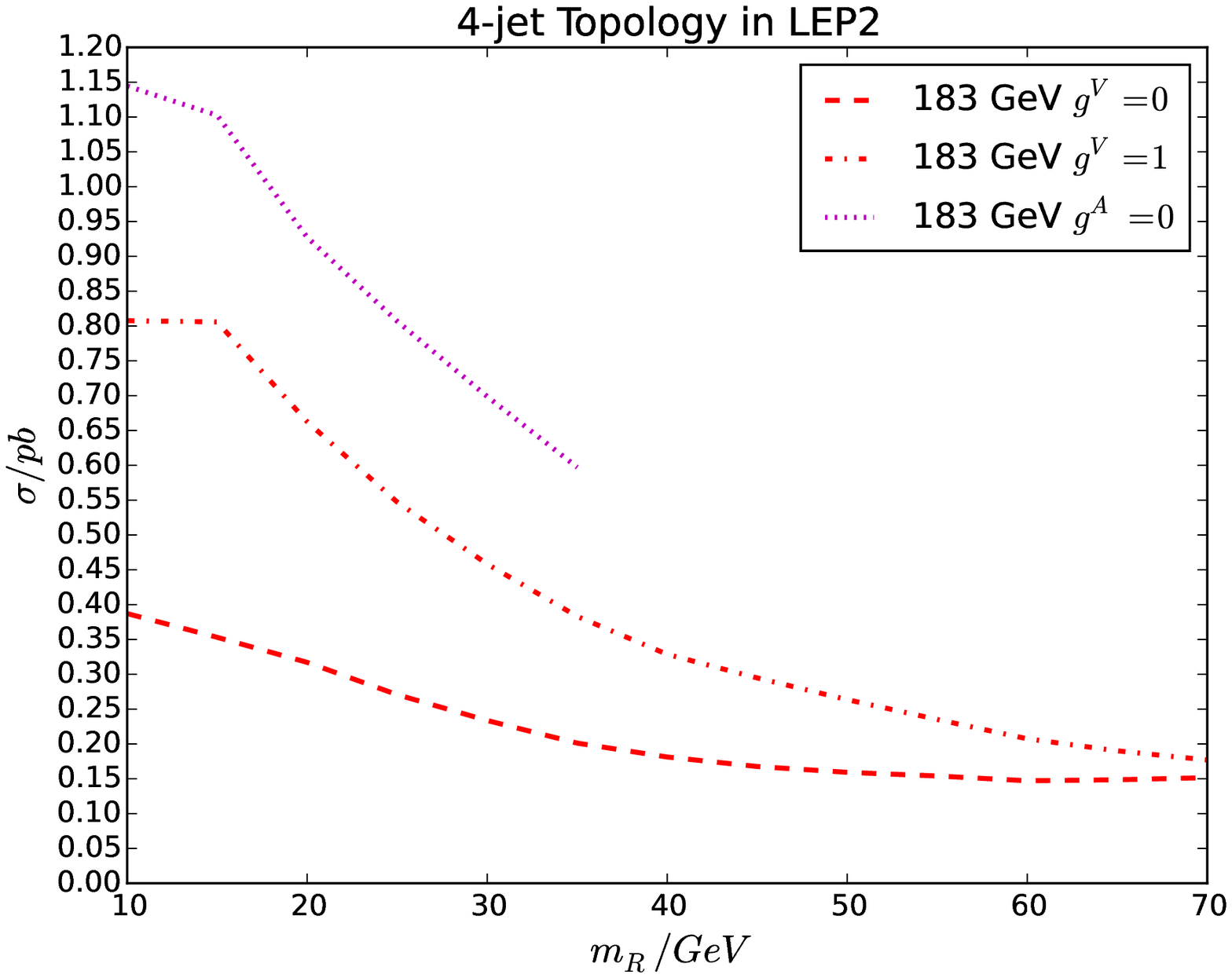}
\caption{Upper bounds on the couplings (left) and total cross section
  (right) from our re--casting of an ALEPH LEP2 $4-$jet analysis. The
  notation is as in Fig.~\ref{fig:05}, except that we only show
  results for the analysis at $\sqrt{s} = 183$ GeV which again has the
  highest sensitivity. For $g^A=0$, shown by the dotted curve, the
  LEP2 bound is only stronger than the perturbativity condition
  (\ref{pert2}) if $m_R<35$ GeV. The unitarity bound on $g^A_b$
  becomes weaker for higher $m_R$, and is no longer relevant. However,
  for $m_\chi > 70$ GeV the perturbativity condition is stronger than
  our ``experimental'' bound, both for $g^V_q=0$ and for $g^V_q=1$.}
\label{fig:06}
\end{figure}

The bounds on the coupling become significantly stronger once
on--shell $R \rightarrow b \bar b$ decays become possible. This region
of larger $m_R$ is explored in Fig.~~\ref{fig:06}. Since the unitarity
bound becomes weaker for higher $m_R$, the final bound on the coupling
is given by our recasting of the LEP2 search until $m_R \simeq 70$ GeV,
where it becomes comparable to the upper bound (\ref{pert2}) from
perturbativity. Over most of the range of $m_R$ shown, the curves for
$g^V_q=0$ and $g^V_q=1$ behave similarly. Nevertheless, there are some
differences for $m_R$ around 10 to 15 GeV. For pure axial vector
coupling the bound on the coupling begins to rise again just after the
point where on--shell $R \rightarrow b \bar b$ decays are allowed.  In
contrast, if the vector coupling is sizable, $g^V_q=1$, the lowest
bound on the axial vector coupling is obtained for $m_R \simeq 12$
GeV. The reason is that contributions due to the exchange of
longitudinal $R$ bosons, which only comes from $g^A$, more strongly
prefer small $m_R$. Hence turning on a vector coupling moves the peak
of the cross section for fixed coupling to slightly larger values of
$m_R$, where on--shell $R \rightarrow b \bar b$ decays are less phase
space suppressed.

We also tried scenarios with pure vector coupling, setting $g^A_b = 0$
(magenta dotted lines in Figs.~\ref{fig:05} and \ref{fig:06}). As in
case of the $2-$jet plus missing energy analysis the resulting bound
on the vector coupling is considerably weaker than that on $g^A$ for
$m_R < 5$ GeV. This is partly due to the absence of the term enhanced
by $m_b^2/m_R^2$, and partly due to the limited cut efficiencies,
which are below $10\%$ in this case. For $m_R < 2 m_b$ the bound on
the vector coupling does not depend much on $m_R$.  The strongest
bound appears for $m_R \gsim 2m_b$. It gradually weakens again for
larger $m_R$, saturating the perturbativity condition (\ref{pert2})
around $m_R=35$ GeV. In this case the cut efficiency is smaller than
in the scenarios with non--vanishing $g^A$, as can be seen from the
larger upper bound on the total signal cross section.

We saw that in case of the two jet plus missing $E_T$ analysis the
strongest bound often comes from LEP1 data rather from LEP2 data, due
to the larger signal cross section (for not too large $m_\chi$) and
smaller SM background. Unfortunately the only published ALEPH new
physics searches at LEP1 using four jet final states
\cite{Decamp:1991uy, Decamp:1991wi} are based on early data samples
with very low integrated luminosity of $1.16$/pb. These searches were
designed for the pair production of particles with sizable coupling to
the $Z$ boson, e.g. light charged Higgs bosons, with each new particle
decaying into a pair of jets. The early data set was sufficient to
exclude such particles with mass nearly up to $M_Z/2$; at the time
there was thus little motivation to perform new physics searches in
the four jet final state using the full LEP1 data sample. Even in the
absence of backgrounds the early analyses could improve on the bounds
we derive from the LEP2 searches only if the cut efficiency of signal
events was higher than $50$\%; this is even higher than the efficiency
of the final states for which these searches were originally
optimized. We conclude that among the published four jet analyses, the
one based on data taken at $\sqrt{s}= 183$ GeV data gives the tightest
constraints on our model.

\section{Improved Analysis of LEP2 Data}
\label{sec:4}

In Sec.~\ref{sec:3} we saw that the cuts applied in the searches for
$2-$jet plus missing energy searches in LEP2 analysis published by the
ALEPH collaboration have limited efficiency for our signal, below
$5$\% in some cases, which is obviously not satisfactory. In this
Section we therefore propose new cuts, which have much better
efficiency for our signal but still remove most SM backgrounds. We
employed kinematical variables also used by ALEPH, and applied them to
SM events generated with Pythia 8.2
\cite{sjostrand2015introduction}. The cuts are designed to remove all
SM backgrounds that may appear for the energy level up to 208
GeV. When analyzing data taken below the $Z$ pair threshold part of
the cuts can be loosened or removed, which would increase the cut
efficiency even further. As the cut efficiency of the $4-$jet
signature as applied to our model is already as good as that of the
$ZH$ search for which this analysis was originally optimized, we only
try to redesign the selection rules for the $jj\slashed{p}$ signature.

We start by requiring at least 5 good tracks in the event; this
essentially guarantees that the event contains some hadronic
activity.

Most multi--hadron events produced by LEP2 were not due to $e^+e^-$
annihilation. Rather, they were produced when both the electron and
the positron emitted a space--like photon; these two photons then
collided to produce a multi--hadronic final state. Since the
virtuality of these photons can be of order of the electron mass, they
can be considered to be quasi--real. This class of events is therefore
often called two--photon (or $\gamma \gamma$) events. The first set of
cuts, listed above the second double line in Table \ref{tab:cuts}, are
designed to greatly reduce this background. These cuts are adapted
from the cuts against two--photon events employed in
ref.\cite{Barate:1997zx}; we use the same kinematical variables, but
change some of the cut values since we optimize our analysis for
higher energies, $\sqrt{s}=208$ GeV.

\begin{table}[htb]
\begin{center}
\begin{tabular}{|c||c|c|}
\hline
\multicolumn{3}{|c|}{Double Jet + Missing Energy ($jj\slashed{p}$)}\\ \hline
Variable & \multicolumn{2}{|c|}{Selection Rule}\\
\hline \hline
$M_{\rm vis}$ & \multicolumn{2}{|c|}{$> 4$ GeV}\\
\hline
$p_T/E_{\rm vis}$ & \multicolumn{2}{|c|}{$> 20\%$}\\
\hline
$\left| \cos \theta_{\rm miss}\right|$ & \multicolumn{2}{|c|}{$< 0.95$}\\
\hline
$\Delta\Phi_T$ & \multicolumn{2}{|c|}{$< 170\degree$}\\
\hline
$E_{\rm NH}$ & \multicolumn{2}{|c|}{$<30\% E_{vis}$} \\ \cline{2-3}
(NH$=$Neutral Hadron) & $30\%\leqslant E_{\rm vis}<45\%$
& $p_{T(\bar{\rm NH})} > 1.8\% \sqrt{s}$\\
\hline
$E_{l_1}/\sqrt{s}$ & \multicolumn{2}{|c|}{$<10\%$}\\
\hline
$E_{12}/\sqrt{s}$ & \multicolumn{2}{|c|}{$<5\%$}\\
\hline \hline
$E_{l_1}^{30}/\sqrt{s}$ & \multicolumn{2}{|c|}{$>1\%$}\\
\hline
Durham & \multicolumn{2}{|c|}{$M_{j_1} \leqslant 9M_{j_2}$ and $M_{j_2}
\leqslant 9M_{j_1}$}\\
\cline{2-3}
2-jet $j_1j_2$ & \multicolumn{2}{|c|}{$M_{j_1j_2}\leqslant 80$ GeV or
$M_{j_1j_2} \geqslant 100$ GeV} \\
\cline{2-3}
$y_{23}$ & \multicolumn{2}{|c|}{$<0.02$}\\ \hline

\end{tabular}
\caption{Cuts designed to reduce the SM background to the two jet plus
  missing $E_T$ signal. The cuts listed above the last horizontal
  double line are mostly directed against two--photon events, whereas
  the cuts below this double line help to remove background events
  containing on--shell $W$ or $Z$ bosons. See the text for further
  details.}
\label{tab:cuts}
\end{center}
\end{table}

The first of these cuts requires the invariant mass of the system
consisting of all detected particles to exceed $4$ GeV. Since the
probability for the emission of a nearly on--shell photon off an
energetic electron or positron quickly increases with decreasing
photon energy, the $\gamma\gamma$ background peaks at small values
of this variable.

The second cut imposes a lower bound on the total transverse momentum
of the system of visible particles. Since photons are mostly emitted
with small virtuality, the $\gamma \gamma$ system typically has small
total transverse momentum. It is nonzero partly due to measurement
errors, and partly because the detector is not hermetic. In
particular, an outgoing $e^{\pm}$ can carry some transverse momentum
but still escape detection. It is important to note that this cut also
removes $e^+e^- \rightarrow q \bar q$ annihilation events (including
events with additional gluon emission), since here the total visible
transverse momentum is also close to zero.

The third cut vetoes events where the missing momentum vector, which is
simply the opposite of the total $3-$momentum of all detected
particles, points nearly in the forward or backward reaction. There is
no reason why the two quasi--real photons should have similar
energy. If the outgoing $e^\pm$ remain undetected, two--photon events
therefore typically have a large longitudinal momentum of detected
particles, i.e. the total missing momentum vector is dominated by its
longitudinal component.

The fourth cut employs the transverse acoplanarity angle
$\Delta\Phi_T$ defined in \cite{heister2003search}. It removes events
where the momenta in two hemispheres are nearly back--to--back. This
cut is again efficient against both two--photon and
$e^+e^- \rightarrow q \bar q$ annihilation events.

The next cut, which we again copy from ALEPH analyses, uses the energy
$E_{\rm NH}$ carried by neutral hadrons. It can be measured by
subtracting the energy associated with the tracks of charged particles
from the total energy measured in the calorimeters. All events where
$E_{\rm NH}$ is less than $30$\% of $E_{\rm vis}$ pass this
cut. Events where $E_{\rm NH} > 0.45 E_{\rm vis}$ are always
removed. If $E_{\rm NH}$ lies between these two values, events only
pass if the visible $p_T$ {\em not} including neutral hadrons, called
$p_{T\bar{\rm NH}}$ in the Table, is at least $0.018 \sqrt{s}$. The
purpose of this combination of cuts is to remove events where a large
fraction of the energy, or of the transverse momentum, is assigned to
neutral hadrons. This can be dangerous, since the energies and momenta
of neutral hadrons are least well determined experimentally of all
``visible'' particles (i.e., not counting neutrinos or DSPs); hence
these events may contain a large amount of ``fake'' missing
(transverse) energy, due to mismeasurement of the neutral hadrons.

The penultimate cut in this category vetoes events with energetic
charged leptons (electrons or muons); $l_1$ is the most energetic identified
charged lepton in the event. This removes two--photon events
where at least one of the photons is so far off--shell that the
corresponding outgoing $e^\pm$ becomes detectable. This cut will also
be effective against other backgrounds, in particular against events
with leptonically decaying $W$ bosons; these events are dangerous
since they also contain a neutrino, which leads to an imbalance of the
visible (transverse) momentum. Of course, events that do not contain
a charged lepton also pass this cut.

The last cut against two--photon events removes events where the
energy $E_{12}$ deposited in forward or backward direction (within
$12\degree$ of the beam axis) exceeds $0.05 \sqrt{s}$. Note that
two--photon events can have a sizable visible energy, even if the
transverse momentum is typically small. This cut also removes events
where one of the outgoing $e^\pm$ hits the detector, but is not
identified as a charged lepton.

The second group of cuts mostly targets events with real $W$ or $Z$
bosons. The first of these uses the variable $E^{30}_{l_1}$, which is
the energy of particles in a $30\degree$ half--angle cone around the
most energetic charged lepton (excluding the lepton itself). This cut
is applied only if the event contains such a lepton. It removes events
where this lepton is isolated, which is typically the case for leptons
from leptonic $W^\pm$ decays. In contrast, charged leptons produced in
the decay of $c$ or $b$ quarks typically have a lot of hadronic activity
nearby, i.e. large values of $E^{30}_{l_1}$, and thus pass this cut.

The three final cuts concern the jet system. In order to apply these
cuts, the event is forced into a two--jet topology using the Durham
$k_T$ algorithm. The first cut removes events where one jet is very
``slim'', i.e. has very small invariant mass. This is often the case
for a jet from a hadronically decaying $\tau$ lepton. This cut thus
removes events containing real $W^\pm \rightarrow \tau^\pm \nu_\tau$
decays. The second cut removes events where the di--jet invariant mass
is close to $M_Z$; this removes $ZZ$ events with one $Z$ boson
decaying hadronically and the other into a neutrino pair,
i.e. invisibly. The last cut removes events where the event would be
reconstructed as containing three or more jets for dimensionless
resolution variable $y_{23} = 0.02$. We find that this cut removes
very efficiently that part of the $e \nu_e W$ background that survived
the lepton cuts.

Some resulting cut efficiencies are listed in Tables~\ref{tab:07} to
\ref{tab:09}. We focus on scenarios with rather light mediator and
light DSP, where the efficiency of our signal for the published
missing energy searches at LEP2, discussed in the previous Section,
was especially poor. For $m_R = 5$ GeV, Table~\ref{tab:07}, we show
efficiencies for pure vector and pure axial vector couplings
separately; for $m_R = 2$ GeV, Table~\ref{tab:08}, and $m_R = 1$ GeV,
Table~\ref{tab:09}, we only show results for pure axial vector
coupling, since outside the region $m_\chi \simeq m_R/2$ the cut
efficiency for pure vector coupling has very little dependence on
$m_R$.

\begin{table}[htb]
\begin{center}
	\begin{tabular}{|c|c|c|c|c|}
		\hline
		\multicolumn{5}{|c|}{$m_R=5$ GeV}\\ \hline
		$m_\chi/$GeV & 2.5 & 3.0 & 3.5 & 4.0\\ \hline
 		$\epsilon_A$ & 18.48\% & 23.53\% & 27.54\% & 29.22\%\\ \hline
 		$\epsilon_V$ & 12.32\% & 15.55\% & 18.09\% & 19.83\%\\ \hline
		$m_\chi/$GeV & 4.5 & 5 & 5.5 & 6.0\\ \hline
		$\epsilon_A$ & 30.74\% & 32.21\% & 33.42\% & 33.58\%\\ \hline
		$\epsilon_V$ & 21.43\% & 22.53\% & 22.97\% & 23.52\%\\ \hline
		$m_\chi/$GeV & 6.5 & 7.0 & 7.5 & 8.0\\ \hline
		$\epsilon_A$ & 34.46\% & 35.02\% & 34.76\% & 35.57\%\\ \hline
		$\epsilon_V$ & 25.94\% & 25.44\% & 26.14\% & 27.25\%\\ \hline
		$m_\chi/$GeV & 8.5 & 9.0 & 9.5 & 10.0\\ \hline
		$\epsilon_A$ & 35.51\% & 35.45\% & 36.15\% & 36.25\%\\ \hline
		$\epsilon_V$ & 27.29\% & 27.26\% & 28.37\% & 29.43\%\\ \hline
	\end{tabular}
 \caption{Cut Efficiencies for $m_R$ = 5 GeV and $m_R/2 \leq m_\chi \leq 2
m_R$. $\epsilon_A$ has been computed with pure axial vector coupling,
$g^V_q = g^V_\chi = 0$, while $\epsilon_V$ is the efficiency for pure
vector coupling, assumed to be the same for $s,c$ and $b$ quarks, while
$g^A_\chi = g^A_q = 0$.}
	\label{tab:07}
\end{center}
\end{table}

\begin{table}[htb]
\begin{center}
	\begin{tabular}{|c|c|c|c|c|}
		\hline
		\multicolumn{5}{|c|}{$m_R=2$ GeV}\\ \hline
		$m_\chi/$GeV & 1.0 & 1.2 & 1.4 & 1.6\\ \hline
 		$\epsilon$ & 21.98\% & 25.82\% & 27.59\% & 29.30\%\\ \hline
		$m_\chi/$GeV & 1.8 & 2.0 & 2.2 & 2.4\\ \hline
		$\epsilon$ & 29.25\% & 30.91\% & 31.08\% & 31.65\%\\ \hline
		$m_\chi/$GeV & 2.6 & 2.8 & 3.0 & 3.2\\ \hline
		$\epsilon$ & 32.07\% & 32.86\% & 33.36\% & 33.65\%\\ \hline
		$m_\chi/$GeV & 3.4 & 3.6 & 3.8 & 4.0\\ \hline
		$\epsilon$ & 33.25\% & 33.40\% & 34.29\% & 34.49\%\\ \hline
	\end{tabular}
 \caption{Cut Efficiencies for $m_R$ = 2 GeV and $m_R/2 \leq m_\chi \leq 2
m_R$. We have assumed pure axial vector coupling, $g^V_q = g^V_\chi = 0$.}
	\label{tab:08}
\end{center}
\end{table}

\begin{table}[htb]
\begin{center}
	\begin{tabular}{|c|c|c|c|c|}
		\hline
		\multicolumn{5}{|c|}{$m_R=1\ GeV$}\\ \hline
		$m_\chi/GeV$ & 0.5 & 0.6 & 0.7 & 0.8\\ \hline
 		$\epsilon$ & 24.36\% & 27.36\% & 28.99\% & 28.57\%\\ \hline
		$m_\chi/GeV$ & 0.9 & 1.0 & 1.1 & 1.2\\ \hline
		$\epsilon$ & 28.77\% & 29.49\% & 30.71\% & 30.23\%\\ \hline
		$m_\chi/GeV$ & 1.3 & 1.4 & 1.5 & 1.6\\ \hline
		$\epsilon$ & 30.85\% & 30.82\% & 31.43\% & 30.22\%\\ \hline
		$m_\chi/GeV$ & 1.7 & 1.8 & 1.9 & 2.0\\ \hline
		$\epsilon$ & 31.98\% & 31.19\% & 32.38\% & 31.25\%\\ \hline
	\end{tabular}
 \caption{Cut Efficiencies for $m_R$ = 1 GeV and $m_R/2 \leq m_\chi \leq 2
m_R$. We have assumed pure axial vector coupling, $g^V_q = g^V_\chi = 0$.}
	\label{tab:09}
\end{center}
\end{table}

We see that the efficiency for pure vector coupling quickly increases
from $m_\chi =m_R /2$ to $m_\chi \simeq m_R$, and then gradually
increase to 35\% for $m_\chi>25$ GeV. These efficiencies are about
three times higher than those for the published analysis discussed in
the previous Section.

Turning to axial vector couplings, the cut efficiency for any
combination $(m_R,m_\chi)$ is again better than the corresponding one
in the published analysis described in the previous Section. For
example, for $m_R = 2 m_\chi$, the efficiency is more than three times
larger. As in case of vector couplings, the cut efficiency quickly
increases when $m_\chi$ is raised from $m_R/2$ to $m_R$; it continues
to increase more slowly for even higher $m_\chi$, reaching slightly
more than $40$\% for $m_\chi > 30$ GeV. Cut efficiencies of 30 to 40\%
are quite typical for many LEP searches.

The selection cuts were chosen to remove most SM backgrounds. We
simulated $\gamma\gamma$ (i.e.,
$e^+ e^- \rightarrow e^+ e^- q \bar q$) events; events with
hadronically decaying $W^+W^-$ or $ZZ$ pairs leading to events with
four hard partons prior to showering; $Z\bar\nu\nu$, $Z l^+ l^-$ and
$Wl\nu_{l}$ events where the gauge boson decays hadronically; and
$e^+ e^- \rightarrow q \bar q$ annihilation events. We include
``purely hadronic'' final states since they can contain heavy $b$ or
$c$ quarks whose semileptonic decays can produce energetic neutrinos,
and hence lead to significant amounts of missing energy. The
$Z \bar \nu \nu$, $Z l^+ l^-$ and $W l \nu_l$ events include
contributions where the lepton pair comes from the decay of a (nearly)
on--shell $Z$ or $W$ boson, but also contributions that only arise at
third order in electroweak couplings. The latter diagrams do not
contribute very much to the total cross sections for these final
state, but populate different regions of phase space.

Our cuts remove more than $99.9$\% of most of these SM
backgrounds. The exceptions are the $Wl\nu_{l}$ and $Z\bar{\nu}\nu$
final states, where $1.05$\% and $5.03$\%, respectively, of all
generated events pass the cuts.  MadGraph finds total cross sections
of $7.34$ pb and $0.33$ pb, respectively, for these two final states,
leading to a total SM background of about $0.1$ pb. Recall that the
upper bounds on the signal cross section we derived in the previous
section, shown in the right frames of
Figs.~\ref{fig:01}--\ref{fig:04}, were $\gsim 0.1$ pb. 

For parameter choices that saturate these earlier bounds, the new cuts
would therefore lead to comparable signal and background cross
sections. Since we cannot apply the new cuts to the actual data, we
cannot quote the resulting bounds, even if the cut efficiencies are
roughly doubled over a broad range of parameters. In order to give
some idea of the expected improvement, we give some sensitivity
limits, i.e. expected bounds (computed under the assumption that the
observed number of events agrees exactly with the SM prediction). To
this end, we use the $p-$value test of the ``null'' hypothesis (SM
only) for a $95\%$ confidence level. For $m_R=5$ GeV with $g^V=0$, the
upper limit on $g^A$ is improved from $1.48$ to $1.39$ at $m_\chi=2.5$
GeV, and the point that LEP data is weaker than unitarity condition
(\ref{unitarity}) moves from $m_\chi=4.5$ GeV to $5$ GeV. The
sensitivity to the vector couplings increases even more. For $m_R=5$
GeV and $m_\chi=2.5$ GeV with $g^A=0$, the bound of $g^V$ is improved
from $1.36$ to $0.99$. For $m_R=5$ GeV and $m_\chi=10$ GeV, the
expected bound on $g^V$ is improved from $2.15$ to $1.88$ with
$g^A=0$, and from $1.64$ to $1.53$ with non-zero $g^A$ reaching
unitarity bound. We repeat that actual bounds can only be derived by
applying our cuts to real data.

Further optimization of the cuts, in order to maximize $S/B$ or
$S/\sqrt{B}$ where $S$ is the signal and $B$ is the background, should
be possible. For example, the (dominant) $W l \nu_l$ background can
be further reduced by slightly reducing the lower end of the excluded
region of the invariant mass of the di--jet system (the penultimate cut
in Table~\ref{tab:cuts}). However, such an optimization should also include
detector effects, which is difficult for us to do reliably. This analysis
nevertheless makes it appear likely that the bounds we derived in the
previous Section, which used published analyses not optimized for this
final state, can be improved significantly.

\section{Summary and Conclusions}
\label{sec:5}

This study derives constraints from published ALEPH searches, based on
data taken at the LEP collider some twenty years ago, on a simplified
dark matter model. The model features a fermionic dark sector particle
(DSP $\chi$) and a spin--1 mediator $R$ which has sizable couplings to
some quarks but not to leptons. A complete model may contain
additional Higgs bosons to generate $m_R$ and/or additional fermions
for anomaly cancellation (see e.g. \cite{Duerr:2014wra}), but the
presence of these particles should not affect our interpretation of
LEP data. This kind of simplified model has of course been analyzed
previously, in particular in connection with LHC data, which impose
severe constraints from ``monojet'' searches if $m_R > 2 m_\chi$, and
from searches for di--jet resonances for heavy $R$. We therefore focus
on rather light mediators, $m_R \lsim 70$ GeV, and always require
$m_R < 2 m_\chi$ so that on--shell $R \rightarrow \chi \bar \chi$
decays are kinematically forbidden. We also impose unitarity and
perturbativity constraints on the parameters of the model.

We consider two different final states. The new physics production of
two jets plus missing energy and momentum, $jj\slashed{p}$, can only
proceed via off--shell $R$ exchange; the signal is thus proportional
to the square of the product of the mediator's coupling to quarks and
to the DSP. In contrast, in our model the production of $4-$jet final
states can occur through real or virtual $R$ exchange, and the signal
depends only on the mediator's coupling to quarks. We used ALEPH data
since this experiment published analyses of both of these final
states, including complete descriptions of the applied cuts and
numbers of surviving SM background events. This allowed us to recast
these analyses; although we did not implement detector effects, these
are likely to be less important for the signal than for the background
(where they can e.g. create missing momentum).

The best bound on the $jj \slashed{p}$ final state from LEP2 data
(taken at $\sqrt{s}$ well above the $Z$ mass) comes from squark
searches.  Somewhat counter--intuitively the resulting bound on the
couplings becomes {\em stronger} for larger $m_\chi$ if $R$ is very
light and axial vector couplings dominate. This is partly because
increasing $m_\chi$ increases the cut efficiency, since it increases
the kinematical lower bound on the missing energy in the event;
however, the main effect is the increase of the contribution from
longitudinal $R$ bosons, whose matrix element scales like
$g^A_\chi g^A_b m_\chi m_b / m_R^2$. However, even though this is the
most promising among several ALEPH searches for this kind of final
state, the cut efficiency for our model is rather low, less than
$20$\%. In particular, for vanishing axial vector couplings the bound
on the vector coupling is worse than that from perturbativity. In
Section~\ref{sec:4} where therefore devised an optimized set of cuts,
which according to our simulation still removes most SM backgrounds,
but has significantly higher efficiency for $q \bar q \chi \bar \chi$
events in our model.

For $m_\chi \lsim 20$ GeV the best bounds nevertheless come from LEP1
data, taken at $\sqrt{s} \simeq M_Z$, well below the $W^+W^-$ and $ZZ$
production thresholds. We found that an ALEPH analysis looking for
$\nu \bar \nu H$ final states, where $H$ is the SM Higgs boson which
is assumed to decay hadronically, uses cuts that have quite a high
efficiency to $q \bar q \chi \bar\chi$ events in our model. For
example, for $m_R = 1$ GeV and $m_\chi \lsim 20$ GeV it requires
$\sqrt{g^A_b g^A_\chi} \leq 0.1$, see Fig.~\ref{fig:09}. However, LEP1
data cannot probe the region $m_\chi \gsim 25$ GeV for couplings that
respect the unitarity and perturbativity constraints. 

Turning to the four jet final state, we found that ALEPH searches for
$ZH$ production in the all--hadronic final state have quite a good cut
efficiency for $q \bar q q' \bar q'$ production via real or virtual $R$
exchange in our model. The resulting bound on the coupling of the mediator
are roughly comparable to those that follow from $jj\slashed{p}$
final states at LEP2, if the DSP is light and the mediator couples
with equal strength to quarks and to the DSP. This search allows to exclude
new parts of parameter space for $m_R \leq 70$ GeV. For somewhat smaller 
$m_R$ we again expect LEP1 data to be considerably more sensitive, due
to the larger signal cross section and reduced background. Unfortunately
the only published ALEPH analysis of four jet final states at LEP1 used
only about $1$\% of the total integrated luminosity. This was sufficient to
exclude the pair production of new particles with masses up to nearly
the beam energy, which was the purpose of this search, but does not allow
to improve the limits we derive from LEP2 data.

In all cases we found that the Dirac structure of the couplings
(vector or axial vector) affects the bounds significantly. This is
partly due to enhanced contributions from longitudinal $R$ exchange,
which are proportional to axial vector couplings. Moreover, the cut
efficiencies often differ, with pure axial vector couplings usually
leading to higher efficiency; the exception is the di--jet plus
missing energy search at LEP1, where for $m_\chi > 10$ GeV vector
couplings lead to a higher cut efficiency.

In summary, we have shown that LEP data should be able to impose
significant new constraints on the parameter space of dark matter
models with a leptophobic spin--1 mediator, if the mass of the
mediator and/or the dark matter particle are in the (tens of) GeV
range and on--shell decays of the mediator into the dark matter
particles are forbidden. While a published LEP1 search for di--jet
plus missing energy final states already has good efficiency for our
model, even the best published analysis of the same final state using
LEP2 data has quite a low efficiency. Conversely, the best LEP2
analysis of four jet final states is already quite useful for our
purposes, but published LEP1 searches for this final state only use a
small fraction of all data.  Improved analyses of LEP data therefore
hold considerable promise to probe new regions of parameter space of
this class of models.

\FloatBarrier

\Acknowledgements

This work was partially supported by the SFB TR33 funded by the Deutsch
Forschungsgemeinschaft, and partially by the by the German ministry for
scientific research (BMBF).

\end{document}